versions, immediately below this notice, is the defect.  %Below the
BIG version macro there is, commented out, a set of macros %for SMALL
versions. If the BIG ones are commented out and the %SMALL ones are
uncommented they will provide you with two TeX pages %for each sheet
of paper in landscape orientation %(similar to the "little" option in
harvmac) which helps to save %paper and trees. Every little counts.
\documentclass[12pt,a4paper]{article} \usepackage{latexsym}
\font\mybb=msbm10 at 12pt
\def\bb#1{\hbox{\mybb#1}}
\def\Z {\bb{Z}}
\def\R {\bb{R}}
\def\II {\bb{I}}
\makeatletter
\@addtoreset{equation}{section}
\makeatother

\begin{document}
\newcommand{\n}{\nu}
\newcommand{\m}{\mu}

% enkele afkortingen speciaal voor dit stuk gebruikt

\renewcommand{\d}{\mbox{$\partial$}}
\newcommand{\bd}{\mbox{$\bar{\partial}$}}
\newcommand{\M}{\mbox{${\cal M}$}}
\newcommand{\F}{\mbox{${\cal F}$}}
\renewcommand{\H}{\mbox{${\cal H}$}}
\renewcommand{\L}{\mbox{${\cal L}$}}
\newcommand{\G}{\mbox{$\hat{G}$}}
\newcommand{\q}{\mbox{${\cal Q}$}}
\newcommand{\tq}{\mbox{$\tilde{\q}$}}

% enkele afkortingen voor griekse dingen

\newcommand{\s}{\mbox{$\sigma$}}
\newcommand{\f}{\mbox{$\phi$}}
\renewcommand{\O}{\mbox{$\Omega$}}
\newcommand{\x}{\mbox{$\chi$}}
\newcommand{\be}{\begin{equation}}
\newcommand{\ee}{\end{equation}}
\renewcommand{\l}{\mbox{$\lambda$}}

\begin{flushright}
\footnotesize
FTUAM-97-4\\
CERN--TH/97--77\\
{\bf hep-th/9705094}\\
May $14$th, $1997$
\normalsize
\end{flushright}

\par

\begin{center}
{\large\bf Transformation of Black-Hole Hair under Duality 
and Supersymmetry}

\vspace{.6cm}

{\bf
Enrique Alvarez${}^{\clubsuit,\,\,\, \diamondsuit,}$
\footnote{E-mail: {\tt enrial@daniel.ft.uam.es}},
Patrick Meessen${}^{\diamondsuit,}$
\footnote{E-mail: {\tt meessen@delta.ft.uam.es}}
and  Tom\'as Ort\'{\i}n${}^{\spadesuit,}$
\footnote{E-mail: {\tt Tomas.Ortin@cern.ch}}${}^{,}$
\footnote{Address after October 1st: {\it IMAFF, CSIC, 
Calle de Serrano 121, E-28006-Madrid, Spain}}
}
\vskip 0.5cm

${}^{\clubsuit}${\it Instituto de F\'{\i}sica Te\'orica, C-XVI,
Universidad Aut\'onoma de Madrid \\
28049 Madrid, Spain}

\vskip 0.2cm

${}^{\diamondsuit}${\it Departamento de F\'{\i}sica Te\'orica, C-XI,
Universidad Aut\'onoma de Madrid \\
28049 Madrid, Spain}

\vskip 0.2cm

${}^{\spadesuit}${\it Theory Division,
C.E.R.N., 1211 Geneva 23, Switzerland}

\end{center}

\vspace{.3cm}

\centerline{\bf ABSTRACT}

\vskip .3 cm

\begin{quote}

\footnotesize

We study the transformation under the full String Theory duality group
of the observable charges (including mass, angular momentum, NUT charge,
electric, magnetic and different scalar charges) of four dimensional
point-like objects whose asymptotic behavior constitutes a subclass
closed under duality. 

We find that those charges fall into two complex four-dimensional
representations of the duality group. T~duality (including Buscher's
transformations) has an $O(1,2)$ action on them and S~duality a $U(1)$
action. The generalized Bogomol'nyi bound is an $U(2,2)$-invariant built
out of one representations while the other representation (which
includes the angular momentum) never appear in it. The bound is
manifestly duality-invariant.

Consistency between T~duality and supersymmetry seems to require that
{\it primary scalar hair} is included in the generalized Bogomol'nyi
bound. We also find that all known four-dimensional supersymmetric
massless black holes are the T~duals in the time direction of the usual
massive supersymmetric black holes. Non-extreme massless ``black holes''
can be found as the T~duals of the non-extreme black holes. All of them
have primary scalar hair and naked singularities.

\normalsize

\end{quote}

\begin{flushleft}
\footnotesize
CERN--TH/97--77
\normalsize
\end{flushleft}

\newpage

%%%%%%%%%%%%%%%%%%%%%%%%%%%%%%%%%%%%%%%%%%%%%%%%%%%%%%%%%%%%%%%%%%%%%

\section*{Introduction}

Black-hole physics is probably the only non perturbative problem in
gravity in which non trivial progress has recently been made owing to
the different perspective afforded by ``string dualities'' (for a
recent review see e.g.~\cite{maldacena}).

In a previous work \cite{quev} a systematic analysis was made of the
behavior of asymptotic charges under T~duality (see
e.g.~\cite{gipora,aao} for four-dimensional non-rotating black
holes\footnote{Since some of the objects studied are singular, as
  opposite to black holes with a regular horizon, the name black hole
  will be used in a generalized sense for (usually point-like) objects
  described by asymptotics such that a mass, angular momentum etc.~can
  be assigned to them.}.

The main goal of the present work is to extend their results, by
essentially widening the class of metrics considered both by allowing
a more general asymptotic behavior and by including more non-trivial
fields. This simultaneously widens the subgroup of the duality group
that acts on that class preserving the asymptotic behavior.

Therefore, we will define the asymptotic behavior considered
(``TNbh'') and we will identify the subgroup that preserves it (the
``ADS''). We will find that the charges naturally fit in multiplets
under the action of this subgroup and that the Bogomol'nyi bound can
be written as a natural invariant of this subgroup. This was to be
expected since duality transformations in general respect unbroken
supersymmetries, but, since duality transformations in general
transform conserved charges that appear in the Bogomol'nyi bound into
non-conserved charges (associated to primary scalar hair) that, in
principle, do not, the consistency of the picture will require us to
include those non-conserved charges into the generalized Bogomol'nyi
bound. A by-product of our study will be the identification of the
known supersymmetric massless black holes as the T~duals in the time
direction of the usual supersymmetric massive black holes. These are
our main results.

One of the motivations for this work was to try to constrain the
angular momentum of black holes using duality and supersymmetry in
such a way that the extreme limit would never be surpassed. As it is
well known the striking difference between the black-hole extremality
bound and the supersymmetry (or Bogomol'nyi) bound: although teh
angular momentum appears in the extremality bound, it does not enter
the supersymmetry bound. This difference is even more surprising in
view of the fact that in presence of only NUT charge (that is, for some
stationary, non-static, cases) both bounds still coincide; the NUT
charge squared must simply be added to the first member in the two
bounds \cite{kall}.  On the other hand, it is also known that some
T~duality transformations seem to break spacetime supersymmetry making
it non-manifest \cite{susyt}.  These two facts could perhaps give rise
to an scenario in which extremal Kerr-Newman black holes (which are
not supersymmetric) could be dual to some supersymmetric
configuration. At the level of the supersymmetry bounds one would see
the angular momentum transforming under a non-supersymmetry-preserving
duality transformation into a charge that does appear in the
supersymmetry bound (like the NUT charge). In this way, the
constraints imposed by supersymmetry on the charge would constraint
equally the angular momentum.

Although this scenario has been disproved by the calculations that we
are going to present\footnote{In fact, the angular momentum is part of
  a set of charges which transform amongst themselves under duality
  and never appear in the Bogomol'nyi bound.} the transformation of
black-hole charges and the corresponding Bogomol'nyi bounds under
general string duality transformations remains an interesting subject
in its own right and its study should help us gain more insight into
the physical space-time meaning of duality.

Thus, in the sequel, the transformation of asymptotic observables (as
multipoles of the metric or of other physical fields) of
four-dimensional ``black holes'' under the T~duality and S~duality
groups will be systematically analyzed, from the four-dimensional
effective action point of view\footnote{The transformation of some of
  the charges we are going to consider here was studied previously in
  Refs.~\cite{ort,kall2}. Here we will extend that study to other sets
  of charges.}.

We are going to consider for simplicity a consistent (from the point
of view of the equations of motion and of the supersymmetry
transformations) truncation\footnote{This truncation is also invariant
  under duality transformations in the compact six-dimensional space
  directions.} of the four-dimensional heterotic string effective
action including the metric, dilaton and two-form field plus two
Abelian vector fields. This truncation is, however, rich enough to
contain solutions with $1/4$ of the supersymmetries unbroken
\cite{klopp,ort}.

Since all the configurations we are going to consider are stationary
and axially symmetric, the theory can be reduced to two dimensions
were the T~dualities due to the presence of isometries in four
dimensions become evident. This we will do in
Section~\ref{sec-derivation} getting manifest $O(2,4)$ due to the
presence of the two Abelian vector fields in four dimensions
\cite{ms}. We will also find the S~duality transformations in their
four-dimensional form.

Then, in Section~\ref{sec-asymptotic} we will define the asymptotic
behavior of those fields in the configurations we are interested in:
(charged, rotating) black holes, Taub-NUT metrics etc. which are
stationary and axially-symmetric. This class of asymptotic behavior
will be referred to as {\it TNbh asymptotics}. Any configuration in
this class will be characterized by a set of parameters ({\it
  charges}) such as the electric and magnetic charges with respect to
the gauge fields, the ADM mass, the angular momentum, the NUT charge
and some other charges forced upon us by duality.

The rest of the paper is organized as follows: In
Section~\ref{sec-transformations} we study the transformation of the
charges under different elements of the T~and S~duality groups and
show explicitly the transformations that preserve TNbh asymptotics
including the effect of constant terms in the asymptotics of the
vector fields (Section~\ref{sec-constant}).  In
Section~\ref{sec-physical} we define the {\it Asymptotic Duality
  Subgroup} as the subgroup of the duality group that preserves TNbh
asymptotics and study the transformation of the Bogomol'nyi bound
under duality. We will find full agreement with the preservation of
unbroken supersymmetry if we admit the presence of primary scalar hair
in the generalized Bogomol'nyi bound. We illustrate this with several
examples in Section~\ref{sec-massless} where we also find that the
known massless supersymmetric black holes are the T~duals of the
common massive supersymmetric ones.  Section~\ref{sec-conclusions}
contains our conclusions.

%%%%%%%%%%%%%%%%%%%%%%%%%%%%%%%%%%%%%%%%%%%%%%%%%%%%%%%%%%%%%%%%%%%%%

\section{The Derivation of the Duality Transformations}
\label{sec-derivation}

In this paper we are going to consider a consistent truncation of the
four-dimensional heterotic string effective action in the string frame
including the metric, axion 2-form and two Abelian vector fields given
by\footnote{Our signature is $(-,+,+,+)$. All hatted symbols are
  four-dimensional and so $\hat{\mu},\hat{\nu}=0,1,2,3$. The relation
  between the four-dimensional Einstein metric
  $\hat{g}_{E\hat{\mu}\hat{\nu}}$ and the string-frame metric
  $\hat{g}_{\hat{\mu}\hat{\nu}}$ is $\hat{g}_{E\hat{\mu}\hat{\nu}}
  =e^{-\hat{\phi}}\hat{g}_{\hat{\mu}\hat{\nu}}$.}:

\begin{equation}
S = \int d^{4}x \sqrt{|\hat{g}|}\ e^{-\hat{\f}} \left[ R(\hat{g})
 +\hat{g}^{\hat{\mu}\hat{\nu}}\d_{\hat{\mu}}\hat{\f}\d_{\hat{\nu}}\hat{\f}
 -{\textstyle\frac{1}{12}} \hat{H}_{\hat{\mu}\hat{\nu}\hat{\rho}}
\hat{H}^{\hat{\mu}\hat{\nu}\hat{\rho}}
-{\textstyle\frac{1}{4}}\hat{F}^{I}{}_{\hat{\mu}\hat{\nu}}
\hat{F}^{I\hat{\mu}\hat{\nu}}
\right]\; ,
\label{s1.1}
\end{equation}

\noindent where  $I=1,2 $ sums over the Abelian gauge fields
$\hat{A}^{I}{}_{\hat{\mu}}$ with standard field strengths

\begin{equation}
\hat{F}^{I}{}_{\hat{\mu}\hat{\nu}}
=2\partial_{[\hat{\mu}}\hat{A}^{I}{}_{\hat{\nu}]}\, .
\end{equation}

\noindent and the two-form field strength is

\begin{equation}
\hat{H}_{\hat{\mu}\hat{\nu}\hat{\rho}}
= 3 \partial_{[\hat{\mu}}\hat{B}_{\hat{\nu}\hat{\rho}]}
-{\textstyle\frac{3}{2}} \hat{F}^{I}{}_{[\hat{\mu}\hat{\nu}}
\hat{A}^{I}{}_{\hat{\rho}]}\, .
\end{equation}

Before proceeding, an explanation of the origin of this action is in
order.  This action can be obtained from the ten-dimensional heterotic
string effective action by first considering only the lowest order in
$\alpha^{\prime}$ terms (so the Yang-Mills fields and $R^{2}$ terms
are consistently excluded), then compactifying the theory on $T^{6}$
to four dimensions following essentially Ref.~\cite{ms} and afterwards
setting to zero all the scalars and identifying the six Kaluza-Klein
vector fields with the six vector fields that come from the
ten-dimensional axion. This last truncation is done in the equations
of motion and it is perfectly consistent with them and with the
supersymmetry transformation rules. The result of this truncation is
the action of $N=4,d=4$ supergravity \cite{kn:CSF} in the string frame
and with the axion 2-form. Setting to zero four of the six vector
fields one gets the above action.

The above action is invariant under Buscher's T~duality transformations in
the six compact directions because these interchange the vector fields
whose origin is the ten-dimensional metric with the the vector fields
whose origin is the ten-dimensional axion and we have identified these
two sets of fields. There are still some trivial T~duality
transformations which correspond to rotations in the internal compact
space. They correspond to global $O(2)$ rotations of the two vector
fields ($O(6)$ rotations in the full $N=4$ supergravity theory).

The reason why we consider two vector fields instead of six or just
one is that the generating solution for black-hole solutions of the
full $N=4,d=4$ theory only needs two non-trivial vector fields.
Starting from this generating solution and performing T~duality
transformations in the compact space and S~duality transformations (to
be described later) which do not change the Einstein metric one can
generate the most general black-hole solution (if the no-hair theorem
holds). Also, the minimal number of vector fields required in this
theory for allowing solutions with $1/4$ of the $N=4$ supersymmetries
unbroken, is two \cite{klopp,ort}.

As announced in the Introduction, it will be assumed that the metric
has a timelike and a spacelike rotational isometry\footnote{The
  action of rotational isometries has fixed points, while
  translational isometries act with no fixed points.}. The former is
physically associated to the stationary (but not static, in general)
character of the spacetime and the other to the axial
symmetry\footnote{The axis corresponds obviously to the set of fixed
  points of the isometry.}. They comute with each other and, thus, one
can find two coordinates, in this case the time $t$ and the angular
variable $\varphi$, adapted to them, such that the background does not
depend on them.  This, then, implies that the theory can be
dimensionally reduced.  Using the standard technique \cite{ss} the
resulting dimensionally reduced, Euclidean, action turns out to be

\begin{eqnarray}
S & = & \int d^{2}x \sqrt{|g|}\ e^{-\f}
\left[  R(g)+g^{\mu\nu}\d_{\mu}\f\d_{\nu}\f
  +{\textstyle\frac{1}{8}}{\rm Tr}\d_{\mu}\M\d^{\mu}\M^{-1}
\right. \nonumber \\
& & \label{s1.2} \\
& &
\left.
\hspace{3cm}
-{\textstyle\frac{1}{4}}W^{i}_{\mu\nu}(\M^{-1})_{ij}W^{j\mu\nu}\right]\; ,
\nonumber
\end{eqnarray}

\noindent Now the spatial indices $\mu\nu=2,3$ for simplicity and we
also have internal indices $\alpha, \beta=0,1$. The two-dimensional
fields are the metric $g_{\mu\nu}$, six vector fields ${\cal
  K}^{i}{}_{\mu} = (K^{(1)\alpha}{}_{\mu}, K^{(2)}{}_{\alpha\mu},
K^{(3)I}{}_{\mu})$ with the standard Abelian field strengths
$W^{i}{}_{\mu\nu}$ ($i=1,\ldots,6$) and a bunch of scalars
$G_{\alpha\beta},\hat{B}_{\alpha\beta},\hat{A}^{I}{}_{\alpha}$ that
appear combined in the $6\times 6$ matrix ${\cal M}_{ij}$. They are
given by

\begin{equation}
\begin{array}{lcllcl}
G_{\alpha\beta} & = & \hat{g}_{\alpha\beta}\, , &
\phi & = & \hat{\phi}  -{\textstyle\frac{1}{2}}\log \mid\det
G_{\alpha\beta}\mid \, , \\
& & & & & \\
K^{(1)\alpha}{}_{\mu} & = &
\hat{g}_{\mu\beta} (G^{-1})^{\beta\alpha}\, , &
C_{\alpha\beta} & = & {\textstyle\frac{1}{2}}
\hat{A}^{I}{}_{\alpha}\hat{A}^{I}{}_{\beta} +\hat{B}_{\alpha\beta}\, , \\
& & & & & \\
g_{\mu\nu} & = & \hat{g}_{\mu\nu}
-K^{(1)\alpha}{}_{\mu}K^{(1)\beta}{}_{\nu}G_{\alpha\beta}\, , &
K^{(3)I}_{\mu} & = &
\hat{A}^{I}{}_{\mu} -\hat{A}^{I}{}_{\alpha} K^{(1)\alpha}{}_{\mu}\, , \\
& & & & & \\
K^{(2)}{}_{\alpha\ \mu} & = & \hat{B}_{\mu\alpha}
+\hat{B}_{\alpha\beta} K^{(1)\beta}{}_{\mu}
+{\textstyle\frac{1}{2}} \hat{A}^{I}{}_{\alpha}K^{(3)I}{}_{\mu}\, ,
\hspace{-1cm} &
& & \\
\end{array}
\end{equation}

\noindent and

\small
\begin{eqnarray}
(\M_{ij}) &=&
\left(
\begin{array}{ccc}
G^{-1}& -G^{-1}C & -G^{-1}A^{T} \\
& & \\
-C^{T}G^{-1} & G+C^{T}G^{-1}C +A^{T}A &
C^{T}{G^{-1}} A^{T} +A^{T} \\
& & \\
-AG^{-1} & AG^{-1}C+A & \II_{2} + AG^{-1} A^{T} \\
\end{array}
\right)\, ,
\label{s1.3}
\end{eqnarray}
\normalsize

\noindent $A$ being the $2\times 2$ matrix with entries
$\hat{A}^{I}{}_{\alpha}$. If $B$ stands for the $2\times 2$ scalar
matrix $(\hat{B}_{\alpha\beta})$, then the $2\times 2$ scalar matrix
$C$ is given by

\begin{equation}
C = {\textstyle\frac{1}{2}} A^{T}A + B\, .
\end{equation}

Any explicit contribution from the three-form automatically vanishes
in two dimensions, which explains why it does not occur in
Eq.~(\ref{s1.2}). On the other hand, the dynamics of a two-dimensional
vector field is trivial\footnote{The equation of motion of a
  two-dimensional vector field implies that the single independent
  field-strength component is a constant.} and this seems to suggest
that we can safely ignore it. However, the correct procedure to
eliminate the vector fields is to solve their equation of motion and
then substitute the solution into the equations of motion of the other
fields. The equations of motion for the vector fields in the action
above tell us that the components of the fields $({\cal
  M}^{-1})_{ij}W^{j}{}_{\mu\nu}$ are constant.  In Ref.~\cite{sen3}
the constants were chosen to be zero by setting the vector fields
themselves to zero, which can be consistently done at the level of the
action.  This is obviously an additional restriction on the
backgrounds considered\footnote{Other choices could lead to
  two-dimensional cosmological terms.}.  This restriction was also
made (in the purely gravitational sector) in the original paper by
Geroch \cite{geroch} and it has been done in all the subsequent
literature on this subject in the form proposed in Refs.~\cite{kinner}
where it was expressed as the requirement that the background possess
``orthogonal insensitivity'', i.e.~it is invariant under
$(t,\varphi)\rightarrow (-t,-\varphi)$.

This restriction is crucial to obtain an infinite-dimensional algebra
of invariances of the equations of motion of the two-dimensional
system. As we are going to explain, though, in this paper we are not
interested in the infinite-dimensional algebra but only in its
zero-mode subalgebra and so we will not impose this restriction.
Nevertheless, all the configurations that we will explicitly consider
will obey it.

%%%%%%%%%%%%%%%%%%%%%%%%%%%%%%%%%%%%%%%%%%%%%%%%%%%%%%%%%%%%%%%%%%%%%%%%%

\subsection{T~Duality Transformations}

The matrix $\M$ satisfies $\M\L\M\L = \II_{6}$, with

\begin{equation}
 \L \equiv
\left(
\begin{array}{ccc}
0& \II_{2} & 0 \\
\II_{2}& 0 & 0 \\
0& 0& \II_{2} \\
\end{array}
\right)\; .
\label{s1.4}
\end{equation}

It can be immediately seen from Eq.~(\ref{s1.2}) that the dimensionally
reduced action, is invariant under the global transformations given by

\begin{equation}
\M \rightarrow \O\ \M\ \O^{T} \; ,
\hspace{1.5cm}
{\cal K}^{i}{}_{\mu} \rightarrow {\O^{i}}_{j}\ {\cal K}^{j}{}_{\mu}\, ,
\label{s1.5}
\end{equation}

\noindent if the transformation matrices $\Omega$ satisfy the identity

\begin{equation}
\Omega\ {\cal L}\ \Omega^{T}={\cal L}\, .
\end{equation}

The matrix ${\cal L}$ given in Eq.~(\ref{s1.4}) can be diagonalized
and put into the form $\eta\ =\ {\rm diag}(-,-,+,+,+,+)$ by a change
of basis associated to the orthogonal matrix $\rho$:

\begin{equation}
\rho\ {\cal L}\ \rho^{T}=\eta\, ,
\hspace{1cm}
 \rho =
\frac{1}{\sqrt{2}}
\left(
\begin{array}{crr}
\II_{2} & -\II_{2} & 0               \\
\II_{2} & \II_{2}  & 0               \\
0       &  0       & \sqrt{2}\II_{2} \\
\end{array}
\right)\, ,
\hspace{1cm}
\rho\ \rho^{T}= \II_{6}\, ,
\end{equation}

\noindent where now $\eta$ is the diagonal metric of $O(2,4;\R)$ which implies
that the $\Omega$'s are $O(2,4;\R)$ transformations in a non-diagonal
basis.  The transformations in the diagonal ($\Omega_{\eta}$) and
non-diagonal basis are related by

\begin{equation}
\label{eq:diag-nondiag}
\Omega_{\eta}= \rho\ \Omega\ \rho^{T}\, ,
\hspace{1cm}
\Omega_{\eta}\ \eta\ \Omega_{\eta}^{T}= \eta\, .
\end{equation}

This symmetry group corresponds to the classical T~duality group.
From the quantum-mechanical point of view, $O(2,4;\R)$ is broken to
$O(2,4;\Z)$ and this group is an exact perturbative symmetry of string
theory.

We must stress at this point that no S~duality transformations are
included in this group. S~duality is a non-local symmetry while
T~duality consists only on local transformations\footnote{At the level
  of the effective action, of course.}. So, where are the S~duality
transformations that were present in four dimensions?

It is well-known \cite{baka,sen3} that this finite symmetry group can
be extended to the infinite algebra $\widehat{o(2,4)}$. The zero-mode
subalgebra corresponds to the algebra $o(2,4;\R)$ of the symmetry we
just described. The S~duality transformations are included in this
algebra as non-local transformations which are not in the zero-mode
subalgebra.

Observe that we could have proceeded in a completely different way: we
could have started by reducing the theory in the time direction to
three dimensions and we could have dualized in three dimensions all
vectors into scalars (as in Ref.~\cite{sen2}). In this way we would
have gotten two scalars from each vector field: one would be the
electrostatic potential $\hat{A}^{I}{}_{t}$ and the other would be the
magnetostatic potential $\tilde{\hat{A}^{I}}{}_{t}$, non-locally
related to the other three components of the vector. In this
three-dimensional theory, S~duality would be realized by local
transformations rotating the electrostatic and magnetostatic
potentials into each other. Further reduction to two dimensions would
give us a different (``dual'') version of the two-dimensional theory
related to the one we have obtained and we are going to study by a
non-local transformation. The dual theory has also a
$\widehat{o(2,4)}_{2}$ invariance but now the S~duality
transformations are in the zero-mode subalgebra $o(2,4;\R)_{2}$
\cite{sen3}.

Another possibility is to study the  S~duality transformations
directly in four dimensions.

%%%%%%%%%%%%%%%%%%%%%%%%%%%%%%%%%%%%%%%%%%%%%%%%%%%%%%%%%%%%%%%%%%%%%

\subsection{S~Duality Transformations}

The $N=4,d=4$ supergravity equations of motion \cite{kn:CSF} have
another duality symmetry nowadays called S~duality that consists of
electric-magnetic duality rotations accompanied of the inversion of
the dilaton (the string coupling constant) and constant shifts of
pseudoscalar axion (see e.g.~Ref.~\cite{sen1} for a review with
references). This symmetry is only manifest in the Einstein frame and
with the pseudoscalar axion. To study it we have to rewrite the action
Eq.~(\ref{s1.1}) in the Einstein frame and then trade the axion
two-form $\hat{B}_{\hat{\mu}\hat{\nu}}$ by the pseudoscalar axion
$\hat{a}$ by means of a Poincar\'e duality transformation.
%
%\begin{equation}
%\label{sdual1}
%\begin{array}{rcl}
%S & = & \int d^{4}x\sqrt{|\hat{g}_{E}|}  \left[ \hat{R}(\hat{g}_{E})
%-{\textstyle\frac{1}{2}}\hat{g}_{E}^{\hat{\mu}\hat{\nu}}
%\partial_{\hat{\mu}}\hat{\phi}\partial_{\hat{\nu}}\hat{\phi}
%-{\textstyle\frac{1}{12}} e^{-2\hat{\phi}}
%\hat{H}_{\hat{\mu}\hat{\nu}\hat{\rho}}\hat{H}^{\hat{\mu}\hat{\nu}\hat{\rho}}
% \right. \\
%& & \\
%& &
%\left. -{\textstyle\frac{1}{4}}e^{-\hat{\phi}}
%\hat{F}^{I}{}_{\hat{\mu}\hat{\nu}} \hat{F}^{I\ \hat{\mu}\hat{\nu}}
%\right] \, . \\
%\end{array}
%\end{equation}
%
(One would get an inconsistent result if one replaced $\hat{H}$ by its
dual field strength directly in the action). Thus, we consider first the
above action as a functional of $\hat{H}$ which is now unrelated to
$\hat{B}$. Then, we have to introduce a Lagrange multiplier ($\hat{a}$)
to enforce the Bianchi identity of $\hat{H}$. Eliminating $\hat{H}$ in
the action by using its equation of motion one finally gets the
following action

\begin{equation}
\label{sdual2}
S= \int d^{4}x\sqrt{|\hat{g}_{E}|} \left[ \hat{R}(\hat{g}_{E})
-{\textstyle\frac{1}{2}}(\partial\hat{\phi})^{2}
-{\textstyle\frac{1}{2}} e^{2\hat{\phi}}(\partial\hat{a})^{2}
-{\textstyle\frac{1}{4}} e^{-\hat{\phi}}\hat{F}^{I}\hat{F}^{I}
+{\textstyle\frac{1}{4}}\hat{a} \hat{F}^{I}{}^{\star}\hat{F}^{I} \right]\, .
\end{equation}

It is important for our purposes to have a very clear relation between
the fields in both formulations since we have to identify the same
charges in both and track them after their transformation. The
(non-local) relation between $\hat{a}$ and $\hat{B}$ and the relation
between the Einstein- and string-frame metric are given by

\begin{equation}
\left\{
\begin{array}{rcl}
\partial_{\hat{\mu}}\hat{a} &  = & \frac{1}{3!\sqrt{|\hat{g}_{E}|}}
e^{-2\hat{\phi}}\hat{\epsilon}_{\hat{\mu}\hat{\nu}\hat{\rho}\hat{\sigma}}
\hat{H}^{\hat{\nu}\hat{\rho}\hat{\sigma}}\, , \\
& & \\
\hat{g}_{E\hat{\mu}\hat{\nu}}
& = &e^{-\hat{\phi}}\ \hat{g}_{\hat{\mu}\hat{\nu}}\, .\\
\end{array}
\right.
\label{eq:pseudoa}
\end{equation}

\noindent Defining now the complex scalar $\hat{\lambda}$ and
the S~dual vector field strengths $\tilde{\hat{F}^{I}}$
\cite{kall2}

\begin{equation}
\hat{\lambda} \equiv \hat{a} +ie^{-\hat{\phi}}\, ,
\hspace{1cm}
\tilde{\hat{F}^{I}}\equiv e^{-\hat{\phi}}\ {}^{\star}\hat{F}^{I} 
+\hat{a}\hat{F}^{I}=\hat{\lambda}\hat{F}^{I\ +} + {\rm c.c.}\, ,
\end{equation}

\noindent where

\begin{equation}
\hat{F}^{I\ \pm}\equiv {\textstyle\frac{1}{2}}\left(\hat{F}^{I}
\mp i\ {}^{\star}\hat{F}^{I} \right)  \, ,
\hspace{1cm}
{}^{\star} \hat{F}^{I\ \pm}= \pm i\hat{F}^{I\ \pm}\, ,
\end{equation}

\noindent  one gets the action

\begin{equation}
\label{sdual3}
S= \int d^{4}x\sqrt{|g_{E}|} \left[ \hat{R}(\hat{g}_{E})
-{\textstyle\frac{1}{2}}\frac{ \partial_{\hat{\mu}}\hat{\lambda}
\partial^{\hat{\mu}}\bar{\hat{\lambda}}}{\left(\Im{\rm m}
\hat{\lambda}\right)^{2}}
+{\textstyle\frac{1}{4}}\hat{F}^{I}\ {}^{\star}\tilde{\hat{F}^{I}} \right]\, .
\end{equation}

The equations of motion plus the Bianchi identities for the vector field
strengths  can be written in the following convenient form

\begin{eqnarray}
\hat{G}_{E\hat{\mu}\hat{\nu}}
+\frac{2}{(\hat{\lambda} -\overline{\hat{\lambda}})}
\left[\partial_{(\hat{\mu}}\hat{\lambda}
\partial_{\hat{\nu})}\overline{\hat{\lambda}}
-{\textstyle\frac{1}{2}}\hat{g}_{E\hat{\mu}\hat{\nu}}
\partial\hat{\lambda}
\partial\overline{\hat{\lambda}}\right]
& & \nonumber \\
& & \nonumber \\
-{\textstyle\frac{1}{4}}
\left(
{}^{\star}\tilde{\hat{F}^{I}}{}_{\hat{\mu}}{}^{\hat{\rho}}
\,\,\,
{}^{\star}\hat{F}^{I}{}_{\hat{\mu}}{}^{\hat{\rho}}
\right)
\left(
\begin{array}{rr}
0 & 1 \\
& \\
-1 & 0 \\
\end{array}
\right)
\left(
\begin{array}{r}
\tilde{\hat{F}^{I}}{}_{\hat{\nu}\hat{\rho}} \\
\\
\hat{F}^{I}{}_{\hat{\nu}\hat{\rho}} \\
\end{array}
\right)
& = & 0\, , \\
& & \nonumber \\
\nabla^{2}\hat{\lambda}
-2 \frac{(\partial\hat{\lambda})^{2}}{(\hat{\lambda}
-\overline{\hat{\lambda}})}
+{\textstyle\frac{i}{8}} (\hat{\lambda} 
-\overline{\hat{\lambda}})^{2}
\left(\hat{F}^{I-}\right)^{2} & = & 0\, , \\
& & \nonumber \\
\nabla_{\hat{\mu}}
\left(
\begin{array}{r}
{}^{\star}\tilde{\hat{F}^{I}\ }{}^{\hat{\mu}\hat{\nu}} \\
\\
{}^{\star}\hat{F}^{I\ \hat{\mu}\hat{\nu}} \\
\end{array}
\right) & = & 0\, .
\end{eqnarray}

In this way, it is easy to see that the last equation is covariant
under linear combinations of the vector fields and the S~dual vector
fields

\begin{equation}
\left(
\begin{array}{r}
\tilde{\hat{F}^{I\prime\ }}{}^{\hat{\mu}\hat{\nu}} \\
\\
\hat{F}^{I\prime\ \hat{\mu}\hat{\nu}} \\
\end{array}
\right)
=
\left(
\begin{array}{cc}
a & b \\
& \\
c & d \\
\end{array}
\right)
\left(
\begin{array}{r}
\tilde{\hat{F}^{I}\ }{}^{\hat{\mu}\hat{\nu}} \\
\\
\hat{F}^{I\ \hat{\mu}\hat{\nu}} \\
\end{array}
\right)\, ,
\label{eq:sdualF}
\end{equation}

\noindent with the only requirement that the transformation matrix is
non-singular. However, these vector fields are not independent and
consistency implies the following non-linear transformations for the
complex scalar $\hat{\lambda}$

\begin{equation}
\label{eq:sduallambda}
\hat{\lambda}^{\prime} = \frac{a\hat{\lambda}+b}{c\hat{\lambda}+d}\, .
\end{equation}

The Einstein equation and the scalar equations are invariant if the
constants $a,b,c,d$ are the entries of an $SL(2,\R)$ ($Sp(2,\R)$)
matrix i.e.

\begin{equation}
ad-bc=1\, .
\end{equation}

These transformations do not act on the Einstein metric. Observe that,
although they are local transformations of the vector field strengths
they are in fact non-local transformations in terms of the true
variables; the vector fields themselves.  Observe that the equations
of motion of the vector fields are nothing but the Bianchi identities
for the dual vector fields
$\tilde{\hat{F}^{I}}{}_{\hat{\mu}\hat{\nu}}$ implying the local
existence of the dual vector fields
$\tilde{\hat{A}^{I}}{}_{\hat{\mu}}$ such that

\begin{equation}
\tilde{\hat{F}^{I}}{}_{\hat{\mu}\hat{\nu}}=
2\partial_{[\hat{\mu}}\tilde{\hat{A}^{I}}{}_{\hat{\nu}]}\, ,
\end{equation}

\noindent which justifies the definition of the $\tilde{\hat{F}^{I}}$'s. 
$\tilde{\hat{A}^{I}}$ depends non-locally on $\hat{A}^{I}$ and the
pair $\tilde{\hat{A}^{I}},\,\,\hat{A}^{I}$ transforms as an $SL(2,\R)$
doublet.

$SL(2,\R)$ is generated by three types of
transformations\footnote{$SL(2,\Z)$ can be generated by the discrete
  versions of the last two.}: rescalings of $\hat{\lambda}$

\begin{equation}
\label{eq:scaling}
\left(
\begin{array}{cc}
a & 0 \\
0 & 1/a \\
\end{array}
\right)\, ,  
\hspace{1cm}
\hat{\lambda}^{\prime} = a^{2}\hat{\lambda}\, ,
\end{equation}

\noindent continuous shifts of the axion

\begin{equation}
\label{eq:axionshifts}
\left(
\begin{array}{rr}
1 & b \\
0 & 1 \\
\end{array}
\right)\, ,  
\hspace{1cm}
\hat{\lambda}^{\prime} = \hat{\lambda} +b\, ,
\end{equation}

\noindent and the discrete transformation

\begin{equation}
\left(
\begin{array}{rr}
0 & 1 \\
-1 & 0 \\
\end{array}
\right)\, ,  
\hspace{1cm}
\hat{\lambda}^{\prime} = -1/\hat{\lambda}\, .
\end{equation}

%%%%%%%%%%%%%%%%%%%%%%%%%%%%%%%%%%%%%%%%%%%%%%%%%%%%%%%%%%%%%%%%%%%%%

\section{TNbh Asymptotics}
\label{sec-asymptotic}

In this section we will present the asymptotic behavior that we will
assume for the solutions of the equations of motion originating from
the action (\ref{s1.1}).

As advertised in the Introduction we are going to consider
generalizations of asymptotically flat Einstein metrics. The
asymptotic behavior of four-dimensional asymptotically flat metrics is
completely characterized to first order in $1/r$ by only two charges:
the ADM mass $M$ and the angular momentum $J$. However, duality
transforms asymptotically flat metrics into non-asymptotically flat
metrics which need different additional charges to be asymptotically
characterized.  One of them \cite{quev} is the NUT charge $N$ and
closure under duality forces us to consider it. We will not need any
more charges in the metric but, for completeness we define a possible
new charge $u$ which we will simply ignore in what follows.

With these conditions on the asymptotics of the four-dimensional metric
it is always possible to choose coordinates such that the
Einstein metric in the $t-\varphi$ subspace has the following
expansion in powers of $1/r$:

\small

\begin{eqnarray}
\label{metricchoice}
\left( \hat{g}_{E\alpha\beta}\right)  \hspace{-.3cm} & = &  \hspace{-.3cm}
 \left(
 \begin{array}{lr}
  -1+2M/r  &
 \hspace{-4cm}2N\cos \theta  +\left[2J\sin^{2}\theta
-4M(N+u)\cos \theta\right]/r\\
& \\
& \\
2N\cos \theta +\left[2J\sin^{2} \theta -4M(N+u)\cos \theta\right]/r &
(r^{2}+2Mr)\sin^{2}\theta
\end{array}
\right)
\nonumber \\
& & \nonumber \\
& & \nonumber \\
& &
+
\left(
\begin{array}{cc}
{\cal O}(r^{-2}) & {\cal O}(r^{-2}) \\
& \\
& \\
{\cal O}(r^{-2}) & {\cal O}(1) \\
\end{array}
\right)\, .\hspace{-3cm}
\end{eqnarray}

\normalsize

We will assume the following behavior for the dilaton

\begin{equation}
e^{-\hat{\phi}} = 1-2{\cal Q}_{d}/r 
+2 {\cal W}\cos\theta/r^{2} -2{\cal Z}/r^{2}
+{\cal O}(r^{-3})\, ,
\end{equation}

\noindent where ${\cal Q}_{d}$ is the dilaton charge,
${\cal W}$ is a charge related to the angular momentum that will be
forced upon us by S~duality and ${\cal Z}$ is a charge which is not
independent, but a function of the electric and magnetic charges (see
below) and is also forced upon us by S~duality .  This implies for the
two-dimensional scalar matrix $G$:

\footnotesize

\begin{eqnarray}
\left( G_{\alpha\beta}\right) \hspace{-.3cm} & = & \hspace{-.3cm}
\left(
\begin{array}{lr}
 -1+2(M-{\cal Q}_{d})/r &
\hspace{-3.8cm}2N\cos \theta  +\left[2J\sin^{2}\theta
-4N(M-{\cal Q}_{d})\cos \theta\right]/r\\
& \\
& \\
2N\cos \theta +\left[2J\sin^{2} \theta
-4N(M-{\cal Q}_{d})\cos \theta\right]/r &
\left[r^{2} +(M+{\cal Q}_{d})r \right]\sin^{2}\theta
\end{array}
\right) \nonumber \\
& & \nonumber \\
& & \nonumber \\
& &
+
\left(
\begin{array}{cc}
{\cal O}(r^{-2}) & {\cal O}(r^{-2}) \\
& \\
& \\
{\cal O}(r^{-2}) & {\cal O}(1) \\
\end{array}
\right)\, .\hspace{-4cm}
\label{s2.1}
\end{eqnarray}

\normalsize

\noindent where we have already set $u=0$.

Observe that we have fixed its constant asymptotic value equal to zero
using the same reasoning as Burgess {\it et.al.}~\cite{quev},
i.e.~rescaling it away any time they arise.  The time coordinate, when
appropriate, will be rescaled as well, in order to bring the
transformed Einstein metric to the above form (i.e.~to preserve our
coordinate (gauge) choice), but in a duality-consistent way.

Sometimes it will also be necessary to rescale the angular coordinate
$\varphi$ in order to get a metric looking like (\ref{metricchoice}).
Conical singularities are then generically induced, and then the
metric is not asymptotically TNbh in spite of looking like
(\ref{metricchoice}).

The objects we will consider will generically carry electric (${\cal
  Q}_{e}^{I}$) and magnetic (${\cal Q}_{m}^{I}$) charges with respect
to the Abelian gauge fields $\hat{A}^{I}{}_{\hat{\mu}}$.  Since we
allow also for angular momentum, they will also have electric (${\cal
  P}_{e}^{I}$) and magnetic (${\cal P}_{m}^{I}$) dipole momenta. This
implies for the two-dimensional scalar matrix $A$ the following
asymptotic behavior

\begin{eqnarray}
\left(\hat{A}^{I}{}_{\alpha}\right) & = &
-2
\left(
\begin{array}{lr}
{\cal Q}^{1}_{e}/r -{\cal P}_{e}^{I}\cos\theta/r^{2}\hspace{1cm}&
{\cal Q}^{1}_{m}\cos \theta +{\cal P}_{m}^{1}\sin^{2}\theta /r \\
& \\
{\cal Q}^{2}_{e}/r -{\cal P}_{e}^{I}\cos\theta/r^{2}&
{\cal Q}^{2}_{m}\cos \theta +{\cal P}_{m}^{2}\sin^{2}\theta /r \\
\end{array}
\right)\nonumber \\
& & \nonumber \\
& & \nonumber \\
& &
+
\left(
\begin{array}{cc}
{\cal O}(r^{-3}) & {\cal O}(r^{-2}) \\
& \\
{\cal O}(r^{-3}) & {\cal O}(r^{-2}) \\
\end{array}
\right)\, .
\label{s2.2}
\end{eqnarray}

Electric dipole momenta appear at higher order in $1/r$ and it is not
strictly necessary to consider them from the point of view of
T~duality, since it will not interchange them with any of the other
charges we are considering and that appear at lower orders in $1/r$.
However, S~duality will interchange the electric and magnetic dipole
momenta and we cannot in general ignore them.

The different behavior of T~and ~duality is due to the fact that
T~duality acts on the potential's components and S~duality acts on the
field strengths.  Thus, for the purpose of performing T~duality
transformations the electric charge and the magnetic momentum terms in
the potentials are of the same order in $1/r$. From the point of view
of S~duality, the electric and magnetic charge terms are of the same
order in $1/r$.

To the matrix $A$ in (\ref{s2.2}) we could have added a constant
$2\times 2$ matrix which would be the constant value of the
$t,\varphi$ components of the vector fields at infinity. Usually these
constants are not considered because they can be removed by a
four-dimensional gauge transformation with gauge parameters depending
linearly on $t$ and $\varphi$.

In \cite{quev} it was claimed those constants (in particular a
constant term in the asymptotic expansion of $\hat{A}^{I}{}_{t}$),
although pure gauge, do have an influence on physical characteristics
of the dual solutions (actually this fact was interpreted there as
evidence against the possibility of performing duality with respect to
isometries with non-compact orbits).

However, a glance at the steps necessary to derive the $O(2,4)$
invariance of the dimensionally reduced theory \cite{ms} immediately
reveals the necessity of not only staying in an adapted coordinate
system, but also that the allowed four-dimensional gauge
transformations are those which correspond to two-dimensional gauge
transformations which are obviously independent of cyclic coordinates
(in this case $t$ and $\varphi$) and keep the matrix $A$ invariant.

In other words: a constant shift in the matrix $A$, is not a symmetry
of the two-dimensional theory but relates two inequivalent
vacua\footnote{In any case, one should not be too dogmatic in this
  issue. After all, we are studying only the massless spectrum of
  four-dimensional string theory and performing dimensional reduction
  to two dimensions disregarding all the massive Kaluza-Klein modes
  which are associated to specific functional dependences on the
  coordinates $t$ and $\varphi$. A full answer on whether $t$- or
  $\varphi$-dependent gauge transformations are allowed and their
  effect on the two-dimensional theory can only be obtained from the
  study of the full theory and it is beyond the scope of the
  effective theory that describes the massless spectrum.}.  The
situation from the point of view of S~duality is not different: the
result of the same classical S~duality transformation
(i.e.~$SL(2,\R)$ transformation) depends on the asymptotic constant
values of the dilaton and axion. These can always be absorbed by
further classical S~duality transformations, but they do not relate
equivalent vacua in general.

Thus, at least from the point of view adopted here, constant terms
are indeed physically meaningful. From the point of view of obtaining
a closed class of solutions under duality they are necessary because
they are generated by duality transformations. Setting the constant
terms to zero is just a specific gauge choice (as much as the
coordinate choice made for the metric is also a coordinate choice).
Duality transformations do not respect these gauge choices.  In the
next section we will study the inclusion of these constant terms in a
consistent way by performing gauge transformations and coordinate
changes in all the fields. However, the transformations with constant
terms become very clumsy and we will consider most transformations on
the configurations we are describing in this section, with zero
constant terms.  Only in Section~\ref{sec-constant}, we will briefly
consider a discrete duality transformation on the most general
configuration.

The two-index form will have the usual charge $\q_{a}$.  Closure under
duality again demands the introduction of a new extra parameter
(``charge'') that we denote by $\F$ and which will play an important
role in what follows.  At the same order in $1/r$ it is possible to
define another charge ${\cal H}$ which is not independent, but a
function of the electric and magnetic charges, as we will show. Its
presence, is required by closure of duality but it transforms as a
dependent charge and it does not play a relevant role. The asymptotic
expansion are, then

\begin{eqnarray}
\left(\hat{B}_{\alpha\beta} \right) & = &
2\left(
\begin{array}{lr}
0 &
\hspace{-4cm}\q_{a}\cos \theta  +\F\sin^{2}\theta/r + \H\cos\theta/r \\
& \\
-\q_{a}\cos \theta  -\F\sin^{2}\theta/r -\H\cos\theta/r &
0 \\
\end{array}
\right)\nonumber \\
& & \nonumber \\
& & \nonumber \\
& &
+
\left(
\begin{array}{cc}
0 & {\cal O}(r^{-2}) \\
& \\
{\cal O}(r^{-2}) & 0 \\
\end{array}
\right)\, .
\label{s2.3}
\end{eqnarray}

As we will show in Section~\ref{sec-mapping}, the necessity of the new
charge ${\cal F}$ becomes clear when looking at discrete duality
subgroups checking that the subgroup's multiplication table is
satisfied. From the physical point of view it is clear that the
presence of angular momentum should induce such a charge.

Observe that we could have added a constant term to
$\hat{B}_{t\varphi}$ as well which could be reabsorbed by
four-dimensional gauge transformations which are not allowed from the
two-dimensional point of view. We choose them to be initially zero as
well for simplicity\footnote{A constant term in $\hat{B}_{t\varphi}$
  implies via duality a constant term in $G_{t\varphi}$ which we have
  also initially set to zero for the same reason. We will consider
  both kinds of constant terms in the next section}.

Now we have to show that the asymptotics that have been assumed on the
gauge fields $\hat{A}^{I}$, and on the two-index field $\hat{B}$
correspond to the gauge-invariant charges that one can define by
looking directly in the asymptotic expansions of the field strengths
$\hat{F}^I$, or $\hat{H}$. The field strengths corresponding to the
above potentials are

\begin{eqnarray}
\hat{F}^{I} & = & 2
\left({\cal Q}^{I}_{e} -2\frac{{\cal P}_{e}^{I}}{r}\cos\theta\right)\ 
\frac{1}{r^{2}}dr\wedge dt 
+ 2{\cal P}_{m}^{I}\sin^{2}\theta\frac{1}{r^{2}}dr\wedge d\varphi
\nonumber \\
& & \nonumber \\
& &
+2 \left({\cal Q}_{m}^{I} -2 \frac{{\cal P}_{m}^{I}}{r}\cos\theta \right)
\sin \theta\ d\theta\wedge d\varphi 
-2{\cal P}_{e}^{I}\sin\theta\frac{1}{r^{2}} d\theta\wedge dt
 \nonumber \\
& & \nonumber \\
& & 
+{\cal O}(r^{-3})\, ,
\label{eq:Ffieldstrength}
\end{eqnarray}

\noindent and 

\begin{eqnarray}
\hat{H} & = & -2\left\{ {\cal Q}_{a}
-\left[ \left({\cal Q}_{e}^{I}{\cal Q}_{m}^{I}+{\cal H} \right)
-2{\cal F}\cos\theta\right]\frac{1}{r}
\right\}\sin\theta\ d\theta\wedge dt\wedge d\varphi
\nonumber \\
& & \nonumber \\
& & -2\left[ {\cal F}\sin^{2}\theta
-\left( {\cal Q}_{e}^{I}{\cal Q}_{m}^{I} -{\cal H} \right)\cos\theta
\right]\frac{1}{r^{2}}\ dr\wedge dt\wedge d\varphi
\nonumber \\
& & \nonumber \\
& & + {\cal O}(r^{-2})\, .
\label{eq:Hfieldstrength0}
\end{eqnarray}

Observe that the effect of taking the Hodge dual of $\hat{F}^{I}$ is
equivalent to replacing $({\cal Q}_{e}^{I}, {\cal P}_{e}^{I})$ by
$({\cal Q}_{m}^{I}, {\cal P}_{m}^{I})$ and $({\cal Q}_{m}^{I}, {\cal
  P}_{m}^{I})$ by $(-{\cal Q}_{e}^{I}, -{\cal P}_{e}^{I})$.

Now we have to identify ${\cal H}$. A convenient way of doing this is
to dualize the three-form field strength to find the asymptotics of
the pseudoscalar axion $\hat{a}$ defined in Eq.~(\ref{eq:pseudoa}).
The partial-differential equation $\partial_{\hat{\mu}}\hat{a}$ for
$\hat{a}$ the consistency condition $\partial_{[\hat{\mu}}
\partial_{\hat{\nu}]} \hat{a}=0$ (which is the Bianchi identity for
$\hat{a}$ and, therefore, the equation of motion for $\hat{B}$) has to
be satisfied and this implies that the combination ${\cal Q}_{e}^{I}
{\cal Q}_{m}^{I} -{\cal H}$ vanishes, so

\begin{equation}
{\cal H} = {\cal Q}_{e}^{I}{\cal Q}_{m}^{I}\, ,
\end{equation}

\noindent and we find

\begin{eqnarray}
\hat{H} & = & -2\left\{ {\cal Q}_{a}
-2{\cal F}\cos\theta\frac{1}{r}
+2{\cal Q}_{e}^{I}{\cal Q}_{m}^{I}\frac{1}{r}
\right\}\sin\theta\ d\theta\wedge dt\wedge d\varphi
\nonumber \\
& & \nonumber \\
& & -2{\cal F}\sin^{2}\theta\frac{1}{r^{2}}
dr\wedge dt\wedge d\varphi
+ {\cal O}(r^{-2})\, .
\label{eq:Hfieldstrength}
\end{eqnarray}

From this expression and (\ref{eq:Ffieldstrength}) we see that all
charges considered have a gauge-invariant meaning. The asymptotic
expansion of the pseudoscalar axion $\hat{a}$ is (allowing for a
constant value at infinity $\hat{a}_{0}$ that we will set to zero in
the initial configuration)

\begin{equation}
\hat{a} = \hat{a}_{0} + 2{\cal Q}_{a}/r -2{\cal F}\cos\theta/r^{2} 
+2{\cal Q}_{e}^{I}{\cal Q}_{m}^{I}/r^{2} +{\cal O}(r^{-3})  \, ,
\end{equation}

\noindent which shows that ${\cal Q}_{a}$ is the standard axion charge 
defined, for instance, in Ref.~\cite{ort}. With the pseudoscalar
axion and the dilaton we find the asymptotic expansion of the complex
scalar $\hat{\lambda}$ (allowing also for a non-vanishing asymptotic
value for the dilaton $\hat{\phi}_{0}$)

\begin{equation}
\hat{\lambda} =   \hat{\lambda}_{0} +2 e^{-\hat{\phi}_{0}}\Upsilon/r 
-2 e^{-\hat{\phi}_{0}}\chi\cos\theta/r^{2} 
+2e^{-\hat{\phi}_{0}}\Theta/r^{2}+{\cal O}(r^{-3})\, , 
\end{equation}

\noindent where

\begin{equation}
\begin{array}{rcl}
\hat{\lambda}_{0} & = & \hat{a}_{0} +ie^{-\hat{\phi}_{0}}\, , \\
& & \\
\Upsilon & = & {\cal Q}_{a} -i{\cal Q}_{d}\, , \\
& & \\
\chi & = & {\cal F} -i {\cal W}\, ,\\
& & \\
\Theta & = & {\cal Q}_{e}^{I}{\cal Q}_{m}^{I} -i{\cal Z}\, .\\
\end{array}
\end{equation}

%%%%%%%%%%%%%%%%%%%%%%%%%%%%%%%%%%%%%%%%%%%%%%%%%%%%%%%%%%%%%%%%%%%%%

\subsection{Inclusion of Constant Terms}

The inclusion of constant terms in the asymptotics of the matrices
$G,A$ and $B$ in a consistent way is trickier than it seems at first
sight. Let us start by discussing the modifications needed to include
constant terms in $A$.

In the presence of constant terms in $A$ one has to be very careful
when identifying the {\it right} axion charges. If we consider the
presence of only constant terms $v^{I}$ in $\hat{A}^{I}{}_{t}$ for
the moment

\small
\begin{eqnarray}
\left(\hat{A}^{I}{}_{\alpha}\right) & = &
\left(
\begin{array}{lr}
v^{1} -2{\cal Q}^{1}_{e}/r +2{\cal P}_{e}^{1}\cos\theta/r^{2}
\hspace{1cm}&
-2{\cal Q}^{1}_{m}\cos \theta -2{\cal P}_{m}^{1}\sin^{2}\theta /r \\
& \\
v^{2} -2{\cal Q}^{2}_{e}/r +2{\cal P}_{e}^{2}\cos\theta/r^{2}&
-2{\cal Q}^{2}_{m}\cos \theta -2{\cal P}_{m}^{2}\sin^{2}\theta /r \\
\end{array}
\right)\nonumber \\
& & \nonumber \\
& & \nonumber \\
& &
+
\left(
\begin{array}{cc}
{\cal O}(r^{-3}) & {\cal O}(r^{-2}) \\
& \\
{\cal O}(r^{-3}) & {\cal O}(r^{-2}) \\
\end{array}
\right)\, .
\end{eqnarray}
\normalsize

\noindent the above expression for the axion field strength changes due
to the Chern-Simons terms to

\begin{eqnarray}
\hat{H} & = & -2\left\{
\left({\cal Q}_{a} -{\textstyle\frac{1}{2}}v^{I}{\cal Q}_{m}^{I} \right)
-2\left( {\cal F} - {\textstyle\frac{1}{2}}v^{I}{\cal P}_{m}^{I} \right)
\cos\theta/r 
\right.
\nonumber \\
& & \nonumber \\
& & 
\left.
+2{\cal Q}_{e}^{I}{\cal Q}_{m}^{I}\frac{1}{r}
\right\}\sin\theta\ d\theta\wedge dt\wedge d\varphi
\nonumber \\
& & \nonumber \\
& & 
-2\left({\cal F} -{\textstyle\frac{1}{2}}v^{I}{\cal P}_{m}^{I} \right)
\sin^{2}\theta/r^{2} dr\wedge dt\wedge d\varphi
+ {\cal O}(r^{-2})\, .
\end{eqnarray}

Now, the {\it right} charges are no longer ${\cal Q}_{a}$ and ${\cal
  F}$ but the combinations ${\cal Q}_{a} -{\textstyle\frac{1}{2}}
v^{I}{\cal Q}_{m}^{I}$ and ${\cal F} -
{\textstyle\frac{1}{2}}v^{I}{\cal P}_{m}^{I}$ that appear in
$\hat{H}$. This really means that in presence of constant terms in
$\hat{A}^{I}{}_{t}$ as above, the asymptotic expansion of $B$ that
gives the right charges as in Eq.~(\ref{eq:Hfieldstrength}), and the
one that on has to use is (setting ${\cal H}=0$)

\begin{equation}
\begin{array}{rcl}
\hat{B}_{t\varphi} & = &
2\left({\cal Q}_{a} +\frac{1}{2}v^{I}{\cal Q}_{m}^{I} \right)\cos \theta  
+2\left({\cal F}  +\frac{1}{2}v^{I}{\cal P}_{m}^{I} \right)\sin^{2}\theta/r \\
& & \\
& & 
+2{\cal Q}_{e}^{I} {\cal Q}_{m}^{I} \cos\theta/r
+ {\cal O}(r^{-2})\, ,
\end{array}
\label{eq:Bcorrected}
\end{equation}

Analogous results would have been obtained by creating the constant
terms via a $t$-dependent gauge transformation of the gauge fields
(which induces, due to the Chern-Simons term present in $\hat{H}$ a
gauge transformation of the two-form field) of and looking for a
gauge-independent definition of the axion charges.  This way of
thinking (i.e.~that the terms arise because of gauge transformations
that take us from the gauge in which we have written the asymptotic
expansion of the potentials and metric in the previous section) is the
most appropriate to study the inclusion of constant terms in $G$ and
$B$. For instance, let us now perform a $\varphi$-dependent gauge
transformation of the gauge fields with parameter $\Lambda^{I}=
w^{I}\varphi$ that induces a constant term in
$\hat{A}^{I}{}_{\varphi}$

\begin{equation}
\delta\hat{A}^{I}{}_{\varphi} = w^{I}\, .
\end{equation}

This transformation induces on the two-form field
Eq.~(\ref{eq:Bcorrected}) (taking into account the constant terms
$v^{I}$) a gauge transformation in $B$. To make the story short, we
will simply say that if we consider a general matrix $A$ with constant
terms

\footnotesize
\begin{eqnarray}
\left(\hat{A}^{I}{}_{\alpha}\right) & = &
\left(
\begin{array}{lr}
v^{1} -2{\cal Q}^{1}_{e}/r +2{\cal P}_{e}^{1}\cos\theta/r^{2}
\hspace{1cm}&
w^{1}-2{\cal Q}^{1}_{m}\cos \theta -2{\cal P}_{m}^{1}\sin^{2}\theta /r \\
& \\
v^{2} -2{\cal Q}^{2}_{e}/r +2{\cal P}_{e}^{2}\cos\theta/r^{2}&
w^{2}-2{\cal Q}^{2}_{m}\cos \theta -2{\cal P}_{m}^{2}\sin^{2}\theta /r \\
\end{array}
\right)\nonumber \\
& & \nonumber \\
& & \nonumber \\
& &
+
\left(
\begin{array}{cc}
{\cal O}(r^{-3}) & {\cal O}(r^{-2}) \\
& \\
{\cal O}(r^{-3}) & {\cal O}(r^{-2}) \\
\end{array}
\right)\, ,
\end{eqnarray}
\normalsize

\noindent we must consider a $B$ matrix of the form (we only write the 
$\hat{B}_{t\varphi}$ entry)

\begin{equation}
\begin{array}{rcl}
\hat{B}_{t\varphi} & = &
x + \frac{1}{2}v^{I}w^{I} 
2\left({\cal Q}_{a} +\frac{1}{2}v^{I}{\cal Q}_{m}^{I} \right)
\cos \theta  -w^{I}{\cal Q}_{e}^{I}/r \\
& & \\
& & 
+2\left({\cal F}  +\frac{1}{2}v^{I}{\cal P}_{m}^{I} \right)
\sin^{2}\theta/r  
+2{\cal Q}_{e}^{I} {\cal Q}_{m}^{I} \cos\theta/r
+{\cal O}(r^{-2})\, , \\
\end{array}
\label{eq:Bcorrected2}
\end{equation}

\noindent to get an axion field strength of the form 
(\ref{eq:Hfieldstrength}) so the constants ${\cal Q}_{a}$ and ${\cal
  F}$ are still the axion charges. Now, a new $t$- or
$\varphi$-dependent gauge transformation of the form $\Lambda= \delta
v^{I}t + \delta w^{I}\varphi$ is reabsorbed in a redefinition of the
constants $x,v^{I},w^{I}$ and does not affect the charges, that keep
their gauge-invariant meaning.  The constant $x$ can also be generated
or absorbed by a gauge transformation of the two-form field and it
does not induce any other changes in the asymptotics of other fields.

Finally, we will see that duality sometimes creates a constant term in
$\hat{g}_{Et\varphi}$. This term can be reabsorbed or induced by a
reparametrization of the time coordinate $t\rightarrow t -q\varphi$.
This transformation changes $G$ and $A$ to

\footnotesize

\begin{eqnarray}
\left( G_{\alpha\beta}\right) \hspace{-.3cm} & = & \hspace{-.3cm}
\left(
\begin{array}{lr}
-1+2(M-{\cal Q}_{d})/r &
\hspace{-5cm}\left(q+2N\cos \theta\right)
\left[1 -2\left(M-{\cal Q}_{d} \right)/r\right] +2J\sin^{2}\theta/r \\
& \\
& \\
\left(q+2N\cos \theta\right)
\left[1 -2\left(M-{\cal Q}_{d} \right)/r\right] +2J\sin^{2}\theta/r &
\left[r^{2} +(M+{\cal Q}_{d})r \right]\sin^{2}\theta
\end{array}
\right) \nonumber \\
& & \nonumber \\
& & \nonumber \\
& &
+
\left(
\begin{array}{cc}
{\cal O}(r^{-2}) & {\cal O}(r^{-2}) \\
& \\
& \\
{\cal O}(r^{-2}) & {\cal O}(1) \\
\end{array}
\right)\, .\hspace{-4cm}
\label{generalmetric}
\end{eqnarray}

\normalsize

\noindent and

\footnotesize
\begin{eqnarray}
\left(\hat{A}^{I}{}_{\alpha}\right) & = &
\left(
\begin{array}{lr}
v^{1} -2{\cal Q}^{1}_{e}/r +2{\cal P}_{e}^{1}\cos\theta/r^{2}\hspace{1cm} &
(w^{1}-qv^{1})-2{\cal Q}^{1}_{m}\cos \theta \\
& \\
&
+2q{\cal Q}_{e}^{1}/r
-2{\cal P}_{m}^{1}\sin^{2}\theta /r \\
& \\
& \\
v^{2} -2{\cal Q}^{2}_{e}/r +2{\cal P}_{e}^{2}\cos\theta/r^{2} &
(w^{2}-qv^{2})-2{\cal Q}^{2}_{m}\cos \theta \\
& \\
&
+2q{\cal Q}_{e}^{2}/r
-2{\cal P}_{m}^{2}\sin^{2}\theta /r \\
\end{array}
\right)\nonumber \\
& & \nonumber \\
& & \nonumber \\
& &
+
\left(
\begin{array}{cc}
{\cal O}(r^{-3}) & {\cal O}(r^{-2}) \\
& \\
& \\
{\cal O}(r^{-3}) & {\cal O}(r^{-2}) \\
\end{array}
\right)\, .
\label{generalvector}
\end{eqnarray}

\normalsize

\noindent Observe that the electric dipole momenta do not appear
in  the right column because they are of higher order.

It is easy to see that there is no need to do further changes
in $B$. Thus, the most general asymptotic expansions that we will
consider are given by the matrices $G$ in Eq.~(\ref{generalmetric})
$A$ in Eq.~(\ref{generalvector}) and $B$ in Eq.~(\ref{eq:Bcorrected2})
which define gauge-invariant charges in the sense that $t$ and
$\varphi$-dependent gauge transformations $\Lambda^{I}=\delta v^{I}t
+(\delta w^{I}-q\delta v^{I})$ and reparametrizations of the form
$t\rightarrow t +\delta q\varphi$ which are the ones that do not take
us out of the Kaluza-Klein ansatz become simple redefinitions of the
constants $v^{I}\rightarrow v^{I}+\delta v^{I}$ etc.~leaving the
charges invariant (which justifies their name).

%%%%%%%%%%%%%%%%%%%%%%%%%%%%%%%%%%%%%%%%%%%%%%%%%%%%%%%%%%%%%%%%%%%%%

The class of asymptotic behavior just described, determined by the
twelve charges

\begin{equation}
\left\{M,J,N,{\cal Q}_{a},{\cal F},{\cal Q}_{d},
{\cal Q}_{e}^{I},{\cal Q}_{m}^{I},{\cal P}_{m}^{I}\right\}
\end{equation}

\noindent  (with or without constant terms in
the matrices $G,B,A$) will be referred to henceforth as {\it TNbh
  asymptotics}.

%%%%%%%%%%%%%%%%%%%%%%%%%%%%%%%%%%%%%%%%%%%%%%%%%%%%%%%%%%%%%%%%%%%%%

\section{Transformation of the Charges under Duality}
\label{sec-transformations}

In this Section we are going to study the transformation of the
charges of asymptotically TNbh configurations under the T~and
S~duality transformations found in Section~\ref{sec-derivation}.

%%%%%%%%%%%%%%%%%%%%%%%%%%%%%%%%%%%%%%%%%%%%%%%%%%%%%%%%%%%%%%%%%%%%%

\subsection{Deriving the Form for the T~Duality Transformation Matrices}

The problem that now will be tackled is how to generate the explicit
transformations of the full $O(2,4,\R)$ classical T~duality group to
find which subgroup maps TNbh asymptotics into TNbh asymptotics.
$O(2,4)$ is a non-compact, non-connected group and our first task is
to elucidate its structure.

It is known that every element from a group $G$, can be written as a
sequence of operators, which are always part of the connected
component containing the identity $G_{0}$ (which is itself a subgroup
of $G$), and elements from the coset $G/G_{0}$. The action of these
elements on any element of $G$ is to take them from a connected part
to a different connected part. This coset is called the {\it
  mapping-class group} $\pi_{0}(G)$.

$O(2,4)$ has four connected pieces: two correspond to matrices with
determinant $+1$ and two to matrices with determinant $-1$. The former
two connected pieces constitute the subgroup $SO(2,4)$ and are related
to the other two by a discrete transformation that generates the group
$O(2,4)/SO(2,4)=\Z_{2}^{(B)}$.  The two connected components of the
subgroup\footnote{All groups $SO(n,m)$ have two connected components
  \cite{helgas}.} $SO(2,4)$ differ by the sign of the $(1,1)$
component of the matrices of the defining representation. The
component with positive sign contains the identity and is the subgroup
$SO^{\uparrow}(2,4)$ and is related to the other connected component
(which is not a subgroup and we denote by $SO^{\downarrow}(2,4)$) by a
discrete transformation that generates another
$\Z_{2}^{(S)}=SO(2,4)/SO^{\uparrow}(2,4)$ subgroup.

Thus, the mapping-class group of $O(2,4)$ is
$O(2,4)/SO^{\uparrow}(2,4) =\Z_{2}^{(B)}\times \Z_{2}^{(S)}$.

We will study it in detail later. Now we are going to concentrate on
describing the duality transformations in the component connected with
the identity $SO^{\uparrow}(2,4)$.

Every element of the connected component of a group can be written as
a sequence of its  one-parameter-subgroups \cite{gilm} and we are going to
study these first.

In our case these are the exponentiated versions of the generators of
the Lie algebra $so(2,4)$, which we write in the covariant form
$M_{ij}$

\begin{equation}
\Omega_{ij}(\alpha_{ij}) = \exp\{-\alpha_{(ij)}M_{(ij)} \}  \, ,
\end{equation}

\noindent  and which satisfy the commutation relations

\begin{equation}
[M_{ij},M_{kl}] = \eta_{il}M_{jk}-\eta_{ik}M_{jl}
+\eta_{jk}M_{il}-\eta_{jl}M_{ik}\; ,
\end{equation}

\noindent where, again, the indices $i,j,k,l=1\ldots,6$ and
$\eta= {\rm diag}(-,-,+,+,+,+)$ is the diagonal metric of $O(2,4)$.

It should be noted that the action of $O(2,4)$ is 6-dimensional,
which means that the group acts through its vector representation on
the matrix $\M_{ij}$ and on the vectors ${\cal K}^{i}{}_{\mu}$.  The
generators of $so(2,4)$ in the vector representation, denoted by
$\Gamma$, are given by

\begin{equation}
\Gamma \left( M_{ij} \right)^{k}{}_{l} =
2\eta_{l[i} \eta^{k}{}_{j]}\, .
\label{s4.1}
\end{equation}

Upon exponentiation of a single generator, one gets a one-parameter
subgroup. In this way to get all the basic one-parameter subgroups of
$SO^{\uparrow}(2,4)$ {\it in the diagonal basis} with metric $\eta$.
Thus, we still need to transform the on-parameter subgroup
transformations to the non-diagonal basis using
Eq.~(\ref{eq:diag-nondiag}) and finally we can study the effect of
these transformations on the fields using Eq.~(\ref{s1.5}).

%%%%%%%%%%%%%%%%%%%%%%%%%%%%%%%%%%%%%%%%%%%%%%%%%%%%%%%%%%%%%%%%%%%%%

\subsection{The One-Parameter Subgroups of the T~Duality Group}

The one-parameter subgroups of $SO^{\uparrow}(2,4)$ are either boosts
involving one of the indices $1,2$ and one of the indices $3,4,5,6$ or
rotations involving the indices $1$ and $2$ or two of the indices
$3,4,5,6$.

Boost matrices are taken to have the form

\begin{equation}
 \O_{\eta}(\mbox{boost}) =
\left(
\begin{array}{cccc}
{\rm ch} & .. & {\rm sh} & .. \\
       ..& .. &  ..      & .. \\
{\rm sh} & .. & {\rm ch} & .. \\
      .. & .. &   ..     & .. \\
\end{array}
\right)\; ,
\end{equation}

\noindent and generate a non-compact $SO^{\uparrow}(1,1)=\R^{+}$
subgroup and every rotation will be  taken to have the form

\begin{equation}
 \O_{\eta}(\mbox{rotation}) =
\left(
\begin{array}{cccc}
\cos   & .. & \sin & .. \\
  ..   & .. & ..   & .. \\
 -\sin & .. & \cos & .. \\
    .. & .. & ..   & .. \\
\end{array}
\right)\; ,
\end{equation}

\noindent and generates a compact $U(1)$ subgroup.
Here the operators will be labelled by the Lie algebra generator that
generates the operators. For instance, we have\footnote{Throughout we
  shall use the abbreviations ${\rm c}=\cos (\alpha_{ij})$, ${\rm
    s}=\sin (\alpha_{ij})$, ${\rm ch}= \cosh (\alpha_{ij})$ and ${\rm
    sh}=\sinh (\alpha_{ij})$, the $ij$ being the indices of the
  transformation.}

\begin{equation}
\Omega_{\eta 13}
\equiv \exp{\{-\alpha_{13}M_{\eta 13}\}}=
\left(
\begin{array}{cccc}
{\rm ch}   & 0 & {\rm sh} &         \\
  0        & 1 & 0        &         \\
{\rm sh}   & 0 & {\rm ch} &         \\
           &   &          & \II_{3} \\
\end{array}
\right)\, ,
\end{equation}

\noindent and in the non-diagonal basis

\begin{equation}
\Omega_{13}=
\left(
\begin{array}{cccc}
{\rm ch} +{\rm sh} & 0 & 0                  &         \\
                   &   &                    &         \\
0                  & 1 & 0                  &         \\
                   &   &                    &         \\
0                  & 0 & {\rm ch} -{\rm sh} &         \\
                   &   &                    &         \\
                   &   &                    & \II_{3} \\
\end{array}
\right)\, ,
\end{equation}

\noindent etc. Therefore it is not difficult to compute the action
of these subgroups on the background fields.  It turns out that only a
few of them (seven, but only five with a non-trivial action and just
three with different actions on the charges) preserve TNbh
asymptotics.

When the transformations that leave TNbh asymptotics intact are known
the exact change in the asymptotic charges can be computed.  The
charges transform linearly. Actually, the T~duality transformations
close on sets of four charges and their effect can be described by
matrices acting on three four-component charge vectors:

\begin{equation}
\vec{M} \equiv
\left(
\begin{array}{c}
M \\ {\cal Q}_{d} \\ {\cal Q}_{e}^{1} \\ {\cal Q}_{e}^{2} \\
\end{array}
\right)\, ,
\hspace{1cm}
\vec{N} \equiv
\left(
\begin{array}{c}
N \\ {\cal Q}_{a} \\ {\cal Q}_{m}^{1} \\ {\cal Q}_{m}^{2} \\
\end{array}
\right)\, ,
\hspace{1cm}
\vec{J} \equiv
\left(
\begin{array}{c}
J \\ {\cal F} \\ {\cal P}_{m}^{1} \\ {\cal P}_{m}^{2} \\
\end{array}
\right)\, ,
\label{eq:chargevectors}
\end{equation}

\noindent which will be referred to, respectively, as
{\it electric, magnetic} and {\it dipole} charge vectors.

There is a fourth charge vector that contains the electric dipole
momenta ${\cal P}_{e}^{I}$, the dilaton dipole-type charge ${\cal W}$
and an unidentified geometrical charge that we denote by $K$

\begin{equation}
\vec{K}\equiv
\left(
\begin{array}{c}
K \\ {\cal W} \\ {\cal P}_{e}^{1} \\ {\cal P}_{e}^{2} \\
\end{array}
\right)\, .
\label{eq:multipletK}
\end{equation}

\noindent The presence of this fourth charge vector is required 
by S~duality, as we will explain later.

For each TNbh duality transformation there is a unique matrix action
on the three vectors. This, for the moment, can be considered merely a
convenient representation of the duality transformations.  It will be
shown later in the paper that the Bogomol'nyi bound can be rewritten
in terms of our multiplets in an exceedingly convenient way.

Let us examine now examine the interesting duality transformations
case by case.

%%%%%%%%%%%%%%%%%%%%%%%%%%%%%%%%%%%%%%%%%%%%%%%%%%%%%%%%%%%%%%%%%%%%%

\subsubsection{$\O_{13}$}

The action of this subgroup is simply equivalent to a rescaling of the
time coordinate and obviously it preserves TNbh asymptotics.  Using
the inverse rescaling to rewrite the metric in the gauge
(\ref{metricchoice}) we find the the action of this duality
transformation is trivial.

%%%%%%%%%%%%%%%%%%%%%%%%%%%%%%%%%%%%%%%%%%%%%%%%%%%%%%%%%%%%%%%%%%%%%

\subsubsection{$\O_{15}$}

This subgroup preserves TNbh asymptotics and our gauge choice for the
metric (\ref{metricchoice}) and for the matrices $A,B$. In particular
it does not generate any constant term in the $A,B,G$ matrices. Thus,
one can proceed to compute the transformation of the charges.  This
transformation is described by the $4\times 4$ symmetric matrix

\begin{equation}
\Omega_{15}^{(4)} \equiv
\left (
\begin{array}{cccc}
\frac{1+{\rm ch}}{2} & \frac{1-{\rm ch}}{2} &
\frac{{\rm sh}}{\sqrt{2}} & 0 \\
& & & \\
\frac{1-{\rm ch}}{2} & \frac{1+{\rm ch}}{2} &
-\frac{{\rm sh}}{\sqrt{2}} & 0 \\
& & & \\
\frac{{\rm sh}}{\sqrt{2}} & -\frac{{\rm sh}}{\sqrt{2}} &
{\rm ch} & 0 \\
& & & \\
0 & 0 & 0 & 1 \\
\end{array}
\right)\, ,
\label{rep.15}
\end{equation}

\noindent so

\begin{equation}
\tilde{\vec{M}} =   \Omega_{15}^{(4)} \vec{M}\, ,
\hspace{1cm}
\tilde{\vec{N}} =   \Omega_{15}^{(4)} \vec{N}\, ,
\hspace{1cm}
\tilde{\vec{J}} =   \Omega_{15}^{(4)} \vec{J}\, .
\end{equation}

%%%%%%%%%%%%%%%%%%%%%%%%%%%%%%%%%%%%%%%%%%%%%%%%%%%%%%%%%%%%%%%%%%%%%

\subsubsection{$\O_{16}$}

The effect of this transformation is identical to the previous one
with the interchange of the labels $I=1$ and $I=2$. This, the matrix
that describes it on the charges is

\begin{equation}
\Omega_{16}^{(4)} \equiv
\left (
\begin{array}{cccc}
\frac{1+{\rm ch}}{2} & \frac{1-{\rm ch}}{2} &
0 & \frac{{\rm sh}}{\sqrt{2}} \\
& & & \\
\frac{1-{\rm ch}}{2} & \frac{1+{\rm ch}}{2} &
0 & -\frac{{\rm sh}}{\sqrt{2}} \\
& & & \\
0 & 0 & 1 & 0 \\
& & & \\
\frac{{\rm sh}}{\sqrt{2}} & -\frac{{\rm sh}}{\sqrt{2}} &
0 & {\rm ch} \\
\end{array}
\right)\, .
\label{rep.16}
\end{equation}

%%%%%%%%%%%%%%%%%%%%%%%%%%%%%%%%%%%%%%%%%%%%%%%%%%%%%%%%%%%%%%%%%%%%%

\subsubsection{$\O_{24}$}

This transformation is analogous to the transformation $\O_{13}$: its
effect is equivalent to a rescaling of the coordinate $\varphi$ that
preserved it periodicity, which is initially fixed to be $2\pi$,
i.e.~all components of fields with indices $\varphi$ are rescaled, but
the coordinate itself is not rescaled. Now, to go back to our
coordinate choice (\ref{metricchoice}) we have to rescale $\varphi$,
changing its periodicity and, thus, introducing conical singularities.
Therefore, this transformation does not preserve TNbh asymptotics.

%%%%%%%%%%%%%%%%%%%%%%%%%%%%%%%%%%%%%%%%%%%%%%%%%%%%%%%%%%%%%%%%%%%%%

\subsubsection{$\O_{35}$}

The result of this transformation is another asymptotically TNbh
metric written in our gauge (\ref{metricchoice}) up to a rescaling of
the time coordinate and up to constant term in the matrix $A$:

\begin{equation}
v^{1}= \sqrt{2} \sin \alpha_{35}\, ,
\end{equation}

\noindent and this has to be taken into account in the definitions
of the axion charges that have to be identified in the transformed
configurations using the expansion of $B$ in
Eq.~(\ref{eq:Bcorrected}).  The rescaling of the time coordinate can
be performed combining $\omega_{35}$ with an $\omega_{13}$
transformation with the right parameter. The result of this
composition is a one-parameter subgroup of transformations that do
preserve TNbh asymptotics and our gauge choice except for the
non-vanishing $v^{1}$.  The effect of this composite transformation on
the charges can be described by the $4\times 4$ symmetric matrix

\begin{equation}
\left(\Omega_{13}\Omega_{35}\right)^{(4)} \equiv
\left (
\begin{array}{cccc}
\frac{{\rm c}+1}{2{\rm c}} & \frac{{\rm c}-1}{2{\rm c}} &
\frac{1}{\sqrt{2}}\frac{{\rm s}}{{\rm c}}  & 0 \\
& & & \\
\frac{{\rm c}-1}{2{\rm c}} & \frac{{\rm c}+1}{2{\rm c}} &
- \frac{1}{\sqrt{2}}\frac{{\rm s}}{{\rm c}}  & 0 \\
& & & \\
\frac{1}{\sqrt{2}}\frac{{\rm s}}{{\rm c}} &
-\frac{1}{\sqrt{2}}\frac{{\rm s}}{{\rm c}} &
\frac{1}{{\rm c}} & 0 \\
& & & \\
0 & 0 & 0 & 1 \\
\end{array}
\right)\, .
\label{rep.35}
\end{equation}

Now, this matrix is exactly the same as $\Omega_{15}^{(4)}$ with the
replacement

\begin{equation}
\cos\alpha_{35} = 1/\cosh\alpha_{15}\, ,
\end{equation}

\noindent and, so, these transformations are identical on the charges.

%%%%%%%%%%%%%%%%%%%%%%%%%%%%%%%%%%%%%%%%%%%%%%%%%%%%%%%%%%%%%%%%%%%%%

\subsubsection{$\O_{36}$}

This transformation is identical to the previous one with the
interchange of the labels $I=1$ and $I=2$. Thus, it also generates a
constant term in the matrix $A$ which has to be taken care of when
identifying the axion charges of the transformed configurations:

\begin{equation}
v^{2}= \sqrt{2} \sin \alpha_{36}\, ,
\end{equation}

Therefore, although it does not preserve TNbh asymptotics, it can be
combined with an $\Omega_{13}$ transformation into a TNbh-preserving
one-parameter subgroup of transformations that can be described by the
action of the $4\times 4$ symmetric matrix

\begin{equation}
\left(\Omega_{13}\Omega_{36}\right)^{(4)} \equiv
\left (
\begin{array}{cccc}
\frac{{\rm c}+1}{2{\rm c}} & \frac{{\rm c}-1}{2{\rm c}} &
0  &\frac{1}{\sqrt{2}}\frac{{\rm s}}{{\rm c}} \\
& & & \\
\frac{{\rm c}-1}{2{\rm c}} & \frac{{\rm c}+1}{2{\rm c}} &
0  & - \frac{1}{\sqrt{2}}\frac{{\rm s}}{{\rm c}} \\
& & & \\
0 & 0 & 1 & 0 \\
& & & \\
\frac{1}{\sqrt{2}}\frac{{\rm s}}{{\rm c}} &
-\frac{1}{\sqrt{2}}\frac{{\rm s}}{{\rm c}} &
0 & \frac{1}{{\rm c}} \\
\end{array}
\right)\, .
\label{rep.36}
\end{equation}

\noindent on the three four-component charge vectors
$\vec{M},\vec{N},\vec{J}$.

%%%%%%%%%%%%%%%%%%%%%%%%%%%%%%%%%%%%%%%%%%%%%%%%%%%%%%%%%%%%%%%%%%%%%

\subsubsection{$\O_{56}$}

This transformation is just the $SO(2)$ subgroup acting on the gauge
fields only and rotates the electric and magnetic charges and the
magnetic dipole momenta.Thus, it can be described by the $4\times 4$
antisymmetric matrix

\begin{equation}
\Omega_{56}^{(4)} \equiv
\left (
\begin{array}{ccc}
\II_{2} &  &  \\
 & {\rm c} & {\rm s} \\
 & -{\rm s} & {\rm c} \\
\end{array}
\right)\, .
\label{rep.56}
\end{equation}

%%%%%%%%%%%%%%%%%%%%%%%%%%%%%%%%%%%%%%%%%%%%%%%%%%%%%%%%%%%%%%%%%%%%%

\subsection{The Mapping-Class Group T~Duality Transformations 
$\Z_{2}^{(B)}$}
\label{sec-mapping}

The last ``elementary'' duality transformations of $O(2,4)$ that we
have to study are those in the coset
$O(2,4)/SO^{\uparrow}(2,4)=\Z_{2}^{(B)}\times\Z_{2}^{(S)}$. What do
these transformations correspond to? In Ref.~\cite{BJO} the simpler
duality groups $O(1,1)$ and $O(1,2)$ where analyzed in detail and it
was found that the subgroup $\Z_{2}^{(S)}$ is essentially generated by
a reflection in all directions in the scalar $\sigma$-model target
space. Here we can do the same and take the generator of
$\Z_{2}^{(S)}$ as the total reflection $-\II_{6}$. In the same
reference it was also found that the subgroup $\Z_{2}^{(B)}$
corresponds essentially to Buscher's duality transformations
\cite{buscher}.

The generator of $\Z_{2}^{(B)}$ is not unique (it is a coset group).
Two obvious choices correspond to the Buscher transformations in the
directions $t$ and $\varphi$. The Buscher transformation in the
direction $\varphi$ does not preserve TNbh asymptotics and so we will
take as generator of $\Z_{2}^{(B)}$ the Buscher transformation in the
direction $t$, that we denote by $\tau$, with matrix

\begin{equation}
\Omega_{\eta}(\tau) =
\left(
\begin{array}{cc}
+1 &           \\
   & - \II_{5} \\
\end{array}
\right)\, .
\end{equation}

Observe that there is no inconsistency in taking one and not the other
as inequivalent representatives because from the point of view of the
TNbh-preserving duality subgroup they are no related: only an
infinite boost ($\Omega_{14}$) will completely rotate $t$ into
$\varphi$ and, in any case, this subgroup does not preserve TNbh
asymptotics itself.

As was said in Section~\ref{sec-asymptotic}, the necessity of
introducing additional ``charges'' like ${\cal F}$ becomes
evident\footnote{To see it in the continuous subgroups one has to study
  the invertibility of the transformations, which is much harder.}
when one studies discrete duality subgroups like $\Z^{(B)}_{2}$.  If
we analyze the $\tau$-transformation explicitly, we see that the
$\tau$-transform of $\hat{g}_{t\varphi}$ is

\begin{equation}
\tilde{\hat{g}}_{t\varphi}
=\frac{1}{2\Omega}\left\{\hat{g}_{tt} \left[\hat{A}^{I}{}_{t}
\hat{A}^{I}{}_{\varphi}-2\hat{B}_{t\varphi}\right]
-\hat{g}_{t\varphi}\hat{A}^{I}{}_{t}\hat{A}^{I}{}_{t} \right\} \, ,
\label{fout1}
\end{equation}

\noindent where $\Omega$ goes asymptotically as

\begin{equation}
\Omega =  1- \frac{4M}{r} + {\cal O}(r^{-2}) \; .
\end{equation}

Looking at the asymptotic behavior of the terms involved, it is easy
to see that to get a contribution to $J$, the initial configuration
has to have a $r^{-1}\sin^{2}\theta $ term in its asymptotic expansion
of the Kalb-Ramond field. This also shows that $J$ transforms into the
new charge ${\cal F}$ since $\tau^{-1}=\tau$.

The effect of $\tau$ on all the charges can be expressed in terms of
the same symmetric $4\times 4$ matrix $\Omega_{\tau}^{(4)}$

\begin{equation}
\Omega_{\tau}^{(4)}=
\left(
\begin{array}{ccc}
0 & 1 &         \\
1 & 0 &         \\
  &   & \II_{2} \\
\end{array}
\right)\, ,
\label{rep.tau}
\end{equation}

\noindent acting on the charge vectors $\vec{M},\vec{N},\vec{J}$.
The involutive property, that on the charges $\tau^{2}=id$ is
immediately apparent.

%%%%%%%%%%%%%%%%%%%%%%%%%%%%%%%%%%%%%%%%%%%%%%%%%%%%%%%%%%%%%%%%%%%%%

A natural worry at this point is whether a combination of
transformations that do not preserve TNbh asymptotics, can result
in a TNbh asymptotics-preserving transformation.

This is a complicated and time-consuming problem that can only be
handled by computational methods for just products of two
transformations. The result of our (CPU-limited) search is negative.

%%%%%%%%%%%%%%%%%%%%%%%%%%%%%%%%%%%%%%%%%%%%%%%%%%%%%%%%%%%%%%%%%%%%%

\subsection{Constant Parts in the Gauge Fields and the  Closed Set of
Asymptotic ``Charges'' Under $\tau$}
\label{sec-constant}

We have to find the way in which the transformations of the charges
change when we include constant terms in the matrices $G,A,B$. To do a
general study would take to much CPU time. Thus, we will only perform
a full check of only the $\tau$ transformation, although the general
picture should become quite clear from our results for $\tau$ and
other general arguments.

First of all, the consistency in the way we have defined charges and
constant terms (which will be referred to as {\it moduli}) implies
that the moduli transform non-linearly amongst themselves and, thus,
their transformations can be studied by setting to zero the charges.
For the $\tau$ transformation this allows us to immediately
get\footnote{Simultaneous rescalings of the dilaton and the time
  coordinate $t$ are necessary to eliminate the constant value of the
  dilaton at infinity and to get an asymptotically TNbh metric.}

\begin{equation}
\left\{
\begin{array}{rclrcl}
\tilde{v}^{I} & = & v^{I}\, , \hspace{1cm}&
\tilde{q} & = & \xi x\, ,\\
& & \\
\tilde{w}^{I} & = & w^{I}\, , &
\tilde{x} & = & \xi^{-1} q\, .\\
\end{array}
\right.
\end{equation}

\noindent where we have used $\xi = (1- \vec{v}^{\ 2}/2)^{-1}$.

Next, we expect the multiplet structure of the duality transformations
to remain valid in the presence of non-trivial moduli. (The multiplets
contain multipole terms of the same order of different fields.) This
can be checked explicitly, but it also allows us to set to zero all
charges except for those in one multiplet and find their
transformation more easily. The result is that we can describe in all
of them the $\tau$ transformation with a unique moduli-dependent
matrix $\Omega^{(4)}_{\tau}(x,q,v,w)$
\begin{equation}
\Omega^{(4)}_{\tau}(x,q,v,w)=
\left(
\begin{array}{cccc}
-\frac{1}{2}\xi\vec{v}^{\ 2} &  \xi                         &
 \xi v^{1} &  \xi v^{2} \\
& & & \\
  \xi                        & -\frac{1}{2}\xi\vec{v}^{\ 2} &
-\xi v^{1}                    & -\xi v^{2}                    \\
& & & \\
-\xi v^{1}            & \xi v^{1}             &
1 +\xi (v^{1})^{2}    & \xi v^{1}v^{2}       \\
& & & \\
-\xi v^{2}            & \xi v^{2}             &
\xi v^{1}v^{2}        & 1+\xi (v^{2})^{2}    \\
\end{array}
\right)\, .
\end{equation}

Observe that this matrix indeed squares to the identity.

What happens to the other transformations in presence of non-trivial
moduli?  The rule is that now the $4\times 4$ matrices
$\Omega_{ij}^{(4)}$ will become moduli-dependent matrices
$\Omega_{ij}^{(4)}(x,q,v,w)$ and the group multiplication table is
satisfied in the following sense:

\begin{equation}
\Omega_{T_{2}}^{(4)}(\tilde{x},\tilde{q},\tilde{v},\tilde{w}) 
\Omega_{T_{1}}^{(4)}(x,q,v,w) =  
\Omega_{T_{2}\cdot T{1}}^{(4)}(x,q,v,w)  \, ,
\end{equation}

\noindent where $(\tilde{x},\tilde{q},tilde{v},\tilde{w})$ are 
the transformed moduli under $T_{1}$. In the case of $\tau$ we had, trivially

\begin{equation}
\Omega^{(4)}_{\tau}(\tilde{x},\tilde{q},\tilde{v},\tilde{w})
\Omega^{(4)}_{\tau}(x,q,v,w) = \II_{4}\, .
\end{equation}

\noindent because $\Omega^{(4)}_{\tau}(x,q,v,w)$ only depends 
on the $v^{I}$ and these are invariant under $\tau$.

%%%%%%%%%%%%%%%%%%%%%%%%%%%%%%%%%%%%%%%%%%%%%%%%%%%%%%%%%%%%%%%%%%%%%

\subsection{Transformation of the Charges under S~Duality}

The transformation of the electric, magnetic, dilaton and axion
charges under S~duality has been previously studied in
Ref.~\cite{ort,kall2}. Here we are considering more charges and we are
choosing initial configurations with vanishing asymptotic values of
the axion and dilaton. In general, S~duality generates non-vanishing
values of these constants and we will remove them by applying further
S~duality transformations.

Let us first see the effect of general classical S~duality
transformations on arbitrary configurations. It is easy to see that
the transformation (\ref{eq:sduallambda}) acts on the asymptotic value
of $\hat{\lambda}$ as follows \cite{ort,kall2}

\begin{equation}
\hat{\lambda}^{\prime}_{0} 
=\frac{a\hat{\lambda}_{0}+b}{c\hat{\lambda}_{0}+d}\, ,
\end{equation}

\noindent and on its complex charge as follows

\begin{equation}
\Upsilon^{\prime} = 
\left(
\frac{c\overline{\hat{\lambda}}_{0} +d}{c\hat{\lambda}_{0} +d}
\right)
\Upsilon\, .
\end{equation}

\noindent The factor multiplying $\Upsilon$ is just a 
$\hat{\lambda}_{0}$-dependent complex phase and, thus the axion and
dilaton charges are simply rotated into one another.  It is also easy
to see that the additional complex charge that we are considering
here $\chi={\cal F} -i{\cal W}$ transforms exactly as $\Upsilon$.

The effect on the electric and magnetic charges is a bit more
difficult to explain because the electric and magnetic charges that
transform in a natural way under S~duality, and which are the ones
conserved in the quantum theory when the Witten effect \cite{witten}
is taken into account, are not the ones we have defined. To be
precise, the equation of motion and the Bianchi identity tell us that
the two charges that are well defined in the quantum theory and obey
the Dirac-Schwinger-Zwanziger quantization condition are

\begin{equation}
\left\{
\begin{array}{rclcl}
q_{e}^{I} & \sim & \int_{S^{2}_{\infty}} \tilde{\hat{F}^{I}} 
& = & e^{-\hat{\phi}_{0}}{\cal Q}_{e}^{I} -\hat{a}_{0}{\cal Q}_{m}^{I}\, , \\
& & \\
q_{m}^{I} & \sim & \int_{S^{2}_{\infty}} \hat{F}^{I}
& = & {\cal Q}_{m}^{I}\, . \\
\end{array}
\right.
\label{eq:chargedefinitions}
\end{equation}

This pair of charges transform under (\ref{eq:sdualF}) as an $SL(2,\R)$ 
doublet

\begin{equation}
\left(
q_{e}^{I\ \prime}\,\,\,\,
q_{m}^{I\ \prime} 
\right)
= 
\left(
q_{e}^{I}\,\,\,\,
q_{m}^{I} 
\right)
\left(
\begin{array}{rr}
a & -c \\
& \\
-b & d \\
\end{array}
\right)  
\, ,
\end{equation}

\noindent which ensures that the DSZ quantization condition, which can be
written for two dyons in the form

\begin{equation}
\left(
q_{e}^{I\ (1)}\,\,\,\,
q_{m}^{I\ (1)} 
\right)
\left(
\begin{array}{rr}
0\,\,\, & 1 \\
& \\
-1 & 0 \\
\end{array}
\right)
\left(
\begin{array}{c}
q_{e}^{I\ (2)}\\
\\
q_{m}^{I\ (2)}\\
\end{array}
\right)
= 
c n\, ,
\hspace{1cm}
n\in \Z\, ,
\end{equation}

\noindent where $c$ is some constant, is S~duality invariant.
From the relation between the charges $\left( q_{e}^{I\ (1)}\,\,\,\,
  q_{m}^{I\ (1)} \right)$ and the charges ${\cal Q}^{I}_{e},{\cal
  Q}_{m}^{I}$ that we are using (\ref{eq:chargedefinitions}) one
readily finds

\begin{equation}
\begin{array}{rcl}
{\cal Q}_{e}^{I\ \prime} & = & (c\hat{a}_{0} +d){\cal Q}_{e}^{I} 
+c e^{-\hat{\phi}_{0}} {\cal Q}_{m}^{I}\, , \\
& & \\
{\cal Q}_{m}^{I\ \prime} & = & -c e^{-\hat{\phi}_{0}} {\cal Q}_{e}^{I}
+(c\hat{a}_{0} +d){\cal Q}_{m}^{I} \, . \\
\end{array}
\end{equation}

It is easy to see that the electric and magnetic dipole momenta
transform in exactly the same fashion.

Now we have to adapt these formulae to our case in which the original
configuration has $\hat{\lambda}_{0}=i$ and in which we want the
transformed configuration to have also $\hat{\lambda}_{0}^{\prime}=i$.
This can be achieved by applying after the general $SL(2,\R)$
transformation, two transformations
(\ref{eq:scaling},\ref{eq:axionshifts}) with the appropriate values of
$a$ and $b$ to absorb the constant values of the axion and dilaton.
This is equivalent to allow only an $SO(2)$ subgroup of $SL(2,\R)$ to
act on the charges. The result, expressed in terms of the entries of
the original $SL(2,\R)$ matrix is

\begin{equation}
\left(
\begin{array}{c}
{\cal Q}_{e}^{I\ \prime} \\
\\
{\cal Q}_{m}^{I\ \prime} \\
\end{array}
\right)
=
\left(
\begin{array}{cc}
\frac{d}{\sqrt{c^{2}+d^{2}}} & \frac{c}{\sqrt{c^{2}+d^{2}}} \\
& \\
\frac{-c}{\sqrt{c^{2}+d^{2}}} & \frac{d}{\sqrt{c^{2}+d^{2}}} \\
\end{array}
\right)
\left(
\begin{array}{c}
{\cal Q}_{e}^{I} \\
\\
{\cal Q}_{m}^{I} \\
\end{array}
\right)\, ,
\end{equation}

\noindent and similarly for the vector of dipole momenta 
$\left({\cal P}_{m}^{I}\, ,{\cal P}_{e}^{I}\right)$ and

\begin{equation}
\left(
\begin{array}{c}
{\cal Q}_{d}^{\prime} \\
\\
{\cal Q}_{a}^{\prime} \\
\end{array}
\right)
=
\left(
\begin{array}{cc}
\frac{d^{2}-c^{2}}{\sqrt{c^{2}+d^{2}}} & \frac{2cd}{\sqrt{c^{2}+d^{2}}} \\
& \\
\frac{-2cd}{\sqrt{c^{2}+d^{2}}} & \frac{d^{2}-c^{2}}{\sqrt{c^{2}+d^{2}}} \\
\end{array}
\right)
\left(
\begin{array}{c}
{\cal Q}_{d} \\
\\
{\cal Q}_{a} \\
\end{array}
\right)\, ,
\end{equation}

\noindent and, analogously for the charge vector
$\left({\cal W}\, ,{\cal F}\right)$. Observe that the last $SO(2)$
transformation matrix is precisely the square of the former.

It is now clear that the multiplet structure that we built for the
T~duality transformations is not respected by S~duality: the last
three components of the ``electric'' multiplet $\hat{M}$ are rotated
into the last three components of the ``magnetic'' multiplet $\vec{N}$
and vice versa. The same happens with the multiplet $\vec{K}$ defined
in Eq.~(\ref{eq:multipletK}), whose last three components are rotated
into those of the multiplet $\vec{J}$ in exactly the same way, and
vice versa (this is the reason why we  introduced $K$ and $\vec{K}$ in
the first place). To respect the T~duality multiplet structure and, at
the same time incorporate the S~duality multiplet structure it is useful
to introduce the complexified multiplets

\begin{equation}
\label{eq:calM}
\vec{\cal M} \equiv \vec{M} +i\vec{N}=
\left( 
\begin{array}{c}
{\cal M} \\ i \Upsilon \\ \Gamma^{1} \\ \Gamma^{2} \\
\end{array}
\right)\, ,  
\end{equation}

\noindent where

\begin{equation}
{\cal M}\equiv M+iN\, ,
\hspace{1cm}
\Gamma^{I} \equiv {\cal Q}_{e}^{I} + i{\cal Q}_{m}^{I}\, ,  
\end{equation}

\noindent and

\begin{equation}
\label{eq:calJ}
\vec{\cal J} \equiv \vec{K} +i\vec{J}=
\left( 
\begin{array}{c}
{\cal J} \\ i \chi \\ \Pi^{1} \\ \Pi^{2} \\
\end{array}
\right)\, ,  
\end{equation}

\noindent where

\begin{equation}
{\cal J}\equiv K+iJ\, ,
\hspace{1cm}
\Pi^{I} \equiv {\cal P}_{e}^{I} + i{\cal P}_{m}^{I}\, .
\end{equation}

These two complex vectors transform under T~duality with exactly the
same $\Omega^{(4)}_{ij}$ matrices as the real vectors and, under the
above S~duality transformations with the complex $\Sigma^{(4)}$
$SO(2)$ matrix

\begin{equation}
\label{eq:sdualmatrix}
\Sigma^{(4)} =
\left(
\begin{array}{cccc}
1 & & & \\
& & & \\
& e^{2i\theta} & & \\
& & & \\
& & e^{i\theta} & \\
& & & \\
& & & e^{i\theta} \\
\end{array}
\right)\, 
\hspace{1cm}\
\theta= {\rm Arg}(d-ic)\, ,
\end{equation}

\noindent so

\begin{equation}
\vec{\cal M}^{\prime} =  \Sigma^{(4)} \vec{\cal M}\, ,
\hspace{1cm}
\vec{\cal J}^{\prime} =  \Sigma^{(4)} \vec{\cal J}\, . 
\end{equation}

%%%%%%%%%%%%%%%%%%%%%%%%%%%%%%%%%%%%%%%%%%%%%%%%%%%%%%%%%%%%%%%%%%%%%

\section{The  Asymptotic Duality Subgroup}
\label{sec-physical}

We define the Asymptotic Duality Subgroup (ADS) as the subgroup of the
full duality group that respects TNbh asymptotics. In the previous
section we have identified several one-parameter subgroups of the
T~duality part of the ADS and we know that the full S~duality group is
a subgroup of the ADS. However these two subgroups do not commute and,
together, generate a large ADS. We proceed to identify it in the next
section and later we will use it to study the invariance of the
Bogomol'nyi bound relevant for the theory we are considering under it.

%%%%%%%%%%%%%%%%%%%%%%%%%%%%%%%%%%%%%%%%%%%%%%%%%%%%%%%%%%%%%%%%%%%%%

\subsection{Identification of the Asymptotic Duality Subgroup}

First, we are going to identify the T~duality subgroup of the ADS.  As
we have seen, from the point of view of its action on the charges it
has only three non-trivial one-parameter subgroups which we take to be
the ones corresponding to the transformations $\Omega_{15}^{(4)},
\Omega_{16}^{(4)}, \Omega_{56}^{(4)}$. To find the group that they
generate we first study the algebra of their infinitesimal generators
$M^{(4)}_{ij}$

\begin{equation}
\Omega^{(4)}_{ij} =\II_{4}-\alpha_{(ij)}M^{(4)}_{(ij)} \, ,
\end{equation}

\noindent which are given by

\begin{equation}
\begin{array}{ccrl}
M^{(4)}_{15} & = & {\textstyle\frac{1}{\sqrt{2}}}
\left(
\begin{array}{rrrr}
0&0&-1&0\\
0&0&1&0\\
-1&1&0&0 \\
0&0&0&0 \\
\end{array}
\right) \, ,
&\hspace{.3cm} 
M^{(4)}_{16}={\textstyle\frac{1}{\sqrt{2}}}
\left(
\begin{array}{rrrr}
0&0&0&-1 \\
0&0&0&1 \\
0&0&0&0 \\
-1&1&0&0 \\
\end{array}
\right)\, , 
\hspace{-2cm}\\
& & & \\
& \\
M^{(4)}_{56} & = &
\left(
\begin{array}{rrrr}
0&0&0&0 \\
0&0&0&0 \\
0&0&0&-1 \\
0&0&1&0
\end{array}
\right)\, .
& \\
\end{array}
\end{equation}

These infinitesimal generators obey the algebra

\begin{equation}
[M^{(4)}_{56},M^{(4)}_{15}]= M^{(4)}_{16}\, ,
\hspace{.2cm}
[M^{(4)}_{56},M^{(4)}_{16}]=  -M^{(4)}_{15}\, ,
\hspace{.2cm}
[M^{(4)}_{15},M^{(4)}_{16}]=  -M^{(4)}_{56}\, .
\end{equation}

A small calculation of the Killing metric then show that on the base
$\{M^{(4)}_{15},M^{(4)}_{16},M^{(4)}_{56}\}$ the metric is diagonal
with entries $\eta^{(3)}={\rm diag} (+,-,-)$ thus proving that the
algebra is $o(1,2)$ and the group generated by the one-parameter
subgroups is $SO^{\uparrow}(1,2)$ and that the T~duality part of the
ADS (taking into account the discrete transformations) is $O(1,2)$.

This raises now the question as to what is the meaning of the
four-component charge vectors. Clearly they transform in the
four-dimensional reducible representation of $O(1,2)$ furnished by the
matrices $\Omega^{(4)}$.  The only representation of this kind is the
direct sum of a singlet and a vector (three-dimensional)
representation of $O(1,2)$, which in turn means that there is a linear
combination of the charges in each charge vector that is invariant
under the full T~duality part of the ADS. It is easy to see that these
combinations are

\begin{equation}
{\textstyle\frac{1}{\sqrt{2}}}\left(M+{\cal Q}_{d}\right)\, ,
\hspace{1cm}  
{\textstyle\frac{1}{\sqrt{2}}}\left(N+ {\cal Q}_{a}\right)\, ,
\hspace{1cm}  
{\textstyle\frac{1}{\sqrt{2}}}\left(J+ {\cal F}\right)\, .
\end{equation}

The triplets over which T~duality acts in the vector representation of
$SO(1,2)$ are

\begin{equation}
\begin{array}{cccccc}
\vec{M}^{(3)}
& = &
\left( 
\begin{array}{c}
{\textstyle\frac{1}{\sqrt{2}}}\left(M-{\cal Q}_{d}\right) \\
{\cal Q}_{e}^{1} \\
{\cal Q}_{e}^{2} \\
\end{array}
\right)\, ,
\hspace{1cm} &
\vec{N}^{(3)}
& = & 
\left( 
\begin{array}{c}
{\textstyle\frac{1}{\sqrt{2}}}\left(N-{\cal Q}_{a}\right) \\
{\cal Q}_{m}^{1} \\
{\cal Q}_{m}^{2} \\
\end{array}
\right)\, , \\
& & & & & \\
\vec{J}^{(3)}
& = & 
\left( 
\begin{array}{c}
{\textstyle\frac{1}{\sqrt{2}}}\left(J-{\cal F}\right) \\
{\cal P}_{m}^{1} \\
{\cal P}_{m}^{2} \\
\end{array}
\right)\, , \hspace{1cm}&
& & \\
\end{array}
\end{equation}

\noindent and, on this representation the generators of the algebra are

\begin{equation}
\begin{array}{ccrl}
M^{(3)}_{15} & = & {\textstyle\frac{1}{\sqrt{2}}}
\left(
\begin{array}{rrr}
0&1&0 \\
1&0&0 \\
0&0&0 \\
\end{array}
\right) \, ,
&\hspace{1cm} 
M^{(3)}_{16}={\textstyle\frac{1}{\sqrt{2}}}
\left(
\begin{array}{rrr}
0&0&1 \\
0&0&0 \\
1&0&0 \\
\end{array}
\right)\, , \\
& & & \\
& \\
M^{(3)}_{56} & = &
\left(
\begin{array}{rrr}
0&0&0 \\
0&0&-1 \\
0&1&0 \\
\end{array}
\right)\, .
& \\
\end{array}
\end{equation}

We remark for future use that the four-dimensional matrices
$\Omega^{(4)}$ of the $1\oplus 3$ representation of $O(1,2)$ respect
the diagonal $O(2,2)$ metric $\eta^{(4)} = {\rm diag}(+,+,-,-)$ and
are also automatically $O(2,2)$ matrices.

%%%%%%%%%%%%%%%%%%%%%%%%%%%%%%%%%%%%%%%%%%%%%%%%%%%%%%%%%%%%%%%%%%%%%

\subsection{The Bogomol'nyi Bound and its Variation}

In $N=4$ supergravity there are two Bogomol'nyi (B) bounds, of the form

\begin{equation}
M^{2}-|Z_{i}|^{2}\geq 0\, ,\hspace{1cm} i = 1,2  \, ,
\end{equation}

\noindent where the $Z_{i}$'s are the complex skew eigenvalues of the
central charge matrix and are combinations of electric and magnetic
charges of the six graviphotons. These two bounds can be combined into
a single bound by multiplying them and then dividing by $M^{2}$. One
gets, then, a {\it generalized} B bound

\begin{equation}
M^{2} +\frac{|Z_{1}Z_{2}|}{M^{2}} -|Z_{1}|^{2}- |Z_{2}|^{2}\geq 0\, .
\end{equation}

In regular black-hole solutions the second term can be identified with
scalar charges of ``secondary'' type. The identification is, actually
(with zero value for the dilaton at infinity)

\begin{equation}
\frac{|Z_{1}Z_{2}|}{M^{2}} = {\cal Q}^{2}_{d} +{\cal Q}^{2}_{a}\, ,
\end{equation}

\noindent and, taking into account the expression of the  central
charges in terms of the ${\cal Q}_{e,m}^{I}$'s one gets the
generalized B bound\footnote{A general expression of the same kind for
  black holes with regular horizons in genera; theories with scalars
  non-minimally coupled to vector fields has been found in
  \cite{kn:GKK}.}  \cite{ort}

\begin{equation}
M^{2} +{\cal Q}_{d}^{2} + {\cal Q}_{a}^{2}
- {\cal Q}_{e}^{I}{\cal Q}_{e}^{I}
-{\cal Q}_{m}^{I}{\cal Q}_{m}^{I}
\geq 0 \, .
\label{bogo0}
\end{equation}

Note however that this bound is valid only for asymptotically flat
spaces (i.e.~with $N=0$).  This problem can however be overcome by the
reasoning of Ref.~\cite{kall} where it was observed that the NUT charge
$N$ does enter in the generalized B bound. With our definitions the B
bound for asymptotically TNbh spaces takes the form

\begin{equation}
 M^{2} +N^{2}
+{\cal Q}_{d}^{2} + {\cal Q}_{a}^{2}
- {\cal Q}_{e}^{I}{\cal Q}_{e}^{I}
-{\cal Q}_{m}^{I}{\cal Q}_{m}^{I}
\geq 0 \, .
\label{bogo1}
\end{equation}

Now we want to study the invariance of this bound under the T~and
S~duality pieces of the ADS that preserves TNbh asymptotics. We will
not make distinctions between primary and secondary scalar charges
since all we are interested in are the transformation rules of the
scalar charges which are the same for primary- or secondary-type
scalar charges.  We will focus on this distinction in the next
section.

Before perform a direct check, let us analyze what we can expect the
result to be. The T~duality piece of the ADS preserves in general
unbroken supersymmetries of the low-energy string effective action:
one can prove that if one solution admits Killing spinors the dual
solution does as well. Equivalent properties can be checked from the
world-sheet point of view. The only instances in which T~duality seems
not to respect unbroken supersymmetries (at least in a {\it manifest}
fashion from the spacetime point of view) is when a Buscher T~duality
transformation is performed with respect to an isometry with fixed
points, like the isometry in the direction $\varphi$ in our
axially-symmetric case \cite{susyt}.  However, this transformation
does not respect TNbh asymptotics and therefore it does not belong to
the ADS.  S~duality is known to always preserve unbroken supersymmetry
\cite{ort} and, thus, we can expect the B bound to be invariant under
the full ADS.

To study the transformation properties of the B bound under the
physical TNbh asymptotics-preserving duality group it is convenient to
use the diagonal metric of $SO(2,2)$ $\eta^{(4)}={\rm diag}\ 
(1,1,-1,-1)$ already introduced at the beginning of this section.
Using this metric and the charge vectors defined in
Eqs.~(\ref{eq:calM},\ref{eq:calJ}) the B bound can be easily rewritten
in this form:

\begin{equation}
\vec{\cal M}^{\dagger} \eta^{(4)}\vec{\cal M}\geq 0\, .
\end{equation}

In this form the B bound of $N=4,d=4$ supergravity is manifestly
$U(2,2)$-invariant. Observe that $U(2,2)\sim O(2,4)$, although it is not
clear if this fact is a mere coincidence or it has a special significance.
The T~duality piece of the ADS is an $O(1,2)$ subgroup of the $O(2,2)$
canonically embedded in $U(2,2)$ and obviously preserves the B bound.
The S~duality piece of the ADS is a $U(1)$ subgroup diagonally embedded
in $U(2,2)$ through the matrices $\Sigma^{(4)}$ defined in
Eq.~(\ref{eq:sdualmatrix}) and obviously preserve the B bound.

The charges in the vector $\vec{\cal J}$ do not appear in the B bound
and neither T~nor S~duality change this fact. It is not possible to
constrain the values of any of the charges it (in particular $J$)  by 
using duality and supersymmetry, as was suggested in the Introduction.

Although we are not going to study the full ADS generated by the
T~duality  and the S~duality pieces, it is clear that there are
transformations in it that rotate the mass $M$ into the NUT charge $N$
and $J$ into $K$: it is enough to perform first a $\tau$ transformation
to interchange the first and second components of the $U(2,2)$ vectors
$\vec{\cal M}$ and $\vec{\cal J}$, then perform an S~duality
transformation that interchanges the real and imaginary parts of the
second component of those vectors and a further $\tau$-transformation
to bring this rotated component back to the first position.

%%%%%%%%%%%%%%%%%%%%%%%%%%%%%%%%%%%%%%%%%%%%%%%%%%%%%%%%%%%%%%%%%%%%%

\subsection{Primary Scalar Hair and Unbroken Supersymmetry}
\label{sec-massless}

So far we have not discussed in detail the physical meaning of the
charges that define TNbh asymptotics. In particular, we have considered
completely unrestricted charges ${\cal Q}_{d}$ and ${\cal Q}_{a}$.

The dilaton charge ${\cal Q}_{d}$, not being protected by a gauge
symmetry, is not a conserved charge. In four dimensions the Kalb-Ramond
two-form is dual to the pseudoscalar axion and the charge ${\cal Q}_{a}$
is just its charge. Again, ${\cal Q}_{a}$ is not a conserved charge.
This may seem contradictory because in the two-form version there is
indeed a gauge symmetry. However, the two-form conserved charge is
actually associated to one-dimensional extended objects, not to the
point-like objects we are considering here. Thus both charges can be
considered non-conserved scalar charges ({\it hair}).

If these scalars were minimally-coupled scalars the standard no-hair
theorems would apply to them and any non-vanishing value of ${\cal
Q}_{d}$ and ${\cal Q}_{a}$ would imply the presence of naked
singularities. The prototype of this kind of singular solution
with non-trivial scalar hair (called {\it primary hair}) is the
one given in Refs.~\cite{kn:JNWALC} for the theory with a massless
scalar and action

\begin{equation}
S= \int d^{4}x \sqrt{|\hat{g}_{E}|}\ \left[\hat{R} (\hat{g}_{E})
+{\textstyle\frac{1}{2}}\partial_{\hat{\mu}}\hat{\phi}
 \partial^{\hat{\mu}}\hat{\phi}\right]\, .
\end{equation}

The solutions take the form

\begin{equation}
\left\{
\begin{array}{rcl}
d\hat{s}^{2}_{E} & = & W^{\frac{M}{r_{0}}-1}Wdt^{2}
-W^{1-\frac{M}{r_{0}}}\left[ W^{-1}dr^{2} +r^{2}d\Omega^{2}\right]\, ,  \\
& & \\
\hat{\phi} & = & \hat{\phi}_{0} -\frac{{\cal Q}_{d}}{r_{0}}\ln W\, ,\\
\end{array}
\right.
\end{equation}

\noindent where

\begin{equation}
\left\{
\begin{array}{rcl}
W & = & 1-2r_{0}/r\, ,\\
& & \\
r_{0}^{2} & = & M^{2} + {\cal Q}_{d}^{2}\, .\\
\end{array}
\right.
\end{equation}

The three fully independent parameters that characterize each solution
are the mass $M$, the scalar charge\footnote{We use the symbol of the
dilaton charge because these solutions (which are written in the
Einstein frame) are also solutions of the equations of motion of the
low-energy string-effective action Eq.~(\ref{s1.1}) with
$\hat{\phi}$ identified with the dilaton.} ${\cal Q}_{d}$ and the value
of the scalar at infinity $\phi_{0}$. Only when ${\cal Q}_{d}=0$ one has
a regular solution (Schwarzschild). In all other cases there is a
singularity at $r=r_{0}$, where the area of 2-spheres of radius $r$
vanishes.

Before continuing with our discussion a couple of remarks should be
made: first, this whole family of solutions belong to the TNbh class
and, second, observe that the above family of solutions includes a
non-trivial {\it massless} solution. Setting $M=0$ above we find

\begin{equation}
\left\{
\begin{array}{rcl}
d\hat{s}^{2}_{E} & = & dt^{2} -dr^{2} -Wr^{2}d\Omega^{2}\, ,  \\
& & \\
\hat{\phi} & = & \hat{\phi}_{0} -\ln W\, ,
\hspace{1cm}
e^{\hat{\phi}-\hat{\phi}_{0}} =W^{-1}\, ,\\
\end{array}
\right.
\label{eq:massless1}
\end{equation}

\noindent with

\begin{equation}
\label{eq:massless2}
W= 1-\frac{2{\cal Q}_{d}}{r}\, .
\end{equation}

In the full low-energy string effective action, the dilaton and the
axion  are non-minimally coupled scalars, though, and the existence of
black-hole solutions with regular horizons in theories with non-minimally
coupled scalars is known  \cite{kn:GGM,kn:GHS}. In these solutions, the
scalar (dilaton) charge is identical to a certain fixed combinations of
the other, conserved, charges:

\begin{equation}
{\cal Q}_{d}\sim \frac{{\cal Q}_{m}^{I}{\cal Q}_{m}^{I}
-{\cal Q}_{e}^{I}{\cal Q}_{2}^{I}}{2M}\, .
\end{equation}

The same is true of the axion charge in solutions with non-trivial axion
hair and regular horizons \cite{kn:STW,ort,kn:KO,kn:BKO}. The axion
charge is in those cases given by

\begin{equation}
{\cal Q}_{a} \sim \frac{{\cal Q}_{e}^{I}{\cal Q}_{m}^{I}}{2M}\, .
\end{equation}

This kind of scalar hair, whose existence does not imply the presence
of naked singularities is called {\it secondary hair}.  It is clear
that the existence of secondary hair does not preclude the existence
of primary hair. In fact, the solutions above can be interpreted in
the framework of string theory with primary but no secondary hair and
there are solutions which have both kinds of hair at the same time
\cite{kn:ALC2}.

Primary scalar hair always seems to imply the presence of naked
singularities, and the no-hair theorem (if it existed such a general
theorem) should probably be called {\it no-primary hair theorem}.

So, what can duality and supersymmetry tell us about primary scalar
hair? At first sight, nothing. In the standard derivations of the
different B bound formulae only conserved electric and magnetic
charges appear and only when all the scalar hair is secondary and
given by the above formulae one can derive the generalized B bounds of
the previous section in which the scalar charges appear.

Nevertheless, let us consider a simple example: Schwarzschild's
solution (given above just by setting ${\cal Q}_{d}=0$). This solution
has no unbroken supersymmetries, which can be understood in terms of
non-saturation of the B bound ($M \geq 0$). A Buscher T~duality
transformation in the time direction belongs to the physical duality
group and should preserve the supersymmetry properties and asymptotic
behavior of the solution and so it should yield a new solution with no
unbroken supersymmetries and TNbh asymptotics. A short calculation
shows that the dual solutions is exactly the massless solution with
primary scalar hair written above in
Eqs.~(\ref{eq:massless1},\ref{eq:massless2})!  It is easy to check
that this solution admits no $N=4$ Killing spinors and so it has no
unbroken supersymmetries\footnote{The dilatino supersymmetry
  transformation rule would be equal to
  $\delta_{\epsilon}\lambda^{I}\sim
  \not\!\partial\hat{\phi}\epsilon^{I}$ which only vanishes for
  $\epsilon^{I}=0$. ($I$ is an $SU(4)$ index here).}. However, the
fact that this solution has no unbroken supersymmetries would not
have been clear from the B bound point of view , had we used the once-standard
form in which primary hair should not added to it, since its mass and
all the other conserved charges are zero, meaning that the bound
 would be trivially saturated.

All that happened in this transformation is that the mass $M$, which
does appear in the B bound has completely transformed in primary
dilaton charge ${\cal Q}_{d}$ which in principle does not.

After our study of the transformation of charges it is clear that to
reconcile these two results one has to admit that the generalized B
bound formula Eq.~(\ref{bogo1}) does apply to all kinds of scalar
charge and not only to the secondary-type one. Only in this way the
invariance of the B bound becomes consistent with the covariance of
the Killing spinor equations.

Although our reasoning is completely clear when we look on specific
solutions one should be able to derive B bounds including primary
scalar charges using a Nester construction based on the supersymmetry
transformation laws of the fermions of the supergravity theory under
consideration. To be able to do this one has to be able to manage more
general boundary conditions including the seemingly unavoidable naked
singularities that primary hair implies.

Although we have kept this discussion strictly four-dimensional it is
easy to generalize these arguments to higher dimensions. In fact,
solutions generalizing the one above to higher ($d$) dimensions can be
straightforwardly found

\begin{equation}
\label{eq:higherALC}
\left\{
\begin{array}{rcl}
ds^{2} & = & -W^{\frac{M}{r_{0}}-1}W dt^{2}
+W^{\frac{1}{d-3}\left(1-\frac{M}{r_{0}}\right)}
\left[ W^{-1} d\rho^{2} +\rho^{2}d\Omega^{2}_{(d-2)}\right]\, , \\
& & \\
\phi & = & \phi_{0}
+\frac{{\cal Q}_{d}}{r_{0}}\ln W\, .
\end{array}
\right.
\end{equation}

\noindent where

\begin{equation}
W = 1 -\frac{2r_{0}}{\rho^{d-3}}\, ,
\end{equation}

\noindent and now

\begin{equation}
r_{0}^{2} = M^{2} + 2\left({\textstyle\frac{d-3}{d-2}}\right){\cal Q}_{d}  \, .
\end{equation}

For ${\cal Q}_{d}=0$ we recover the $d$-dimensional Schwarzschild
solution. In all other cases we have metrics with naked singularities
either at $\rho=0$ or $\rho^{d-3}=2r_{0}$.

A further example can be useful to fix these ideas.

Using our conventions, it is possible to write the stringy RN solution
in the following form:

\begin{equation}
\left\{
\begin{array}{rcl}
d\hat{s}^{2}_{E} & = & -H^{-2}W dt^{2}
+H^{2}\left[W^{-1}dr^{2} +r^{2}d\Omega^{2} \right]\, , \\
& & \\
e^{-\hat{\phi}} & = & H/H=1\, ,\\
& & \\
\hat{A}^{(1)}{}_{t} & = & 2\alpha_{1}
\frac{|Q|}{M-r_{0}} \left(H^{-1} -1\right)\, , \\
& & \\
\hat{A}^{(2)}{}_{\varphi} & = & -2\alpha_{2} |Q|\cos\theta\, , \\
\end{array}
\right.
\end{equation}

\noindent where $H$ and $W$ are (not independent) harmonic functions

\begin{equation}
H = 1 + \frac{M-r_{0}}{r}\, ,
\hspace{1cm}
W = 1 -\frac{2r_{0}}{r}\, ,
\end{equation}

\noindent and the constants are:

\begin{equation}
\alpha_{i}^{2}=1\, ,
\hspace{1cm}
r_{0}^{2} = M^{2} -2Q^{2}\, ,
\end{equation}

\noindent where  we have set

\begin{equation}
{\cal Q}_{e}^{1}= \alpha_{1}|Q|\, ,
\hspace{1cm}
 {\cal Q}_{m}^{2}= \alpha_{2} |Q|\, .
\end{equation}

The dilaton charge is identically zero for this family.  Observe also
that $M-r_{0}\geq 0$ always, and thus $H$ never vanishes and so it
never gives rise to any singularities in the metric apart from the one
at $r=0$, which is the curvature singularity. The metric is also
singular at the horizon $r=2r_{0}>0$ where $W$ vanishes, covering the
physical singularity at $r=0$.

The extremal limit is $r_{0}=0$, $M=\sqrt{2}|Q|$, which makes $W$
disappear and $H$ becomes an unrestricted harmonic function (we could
describe many BHs if we wanted). In this limit the horizon is placed at
$r=0$, which is th locus of a two-sphere instead of a point, as can be
seen by a coordinate change. The curvature singularity is not covered by
these coordinates.

Th B bound for this family of solutions is

\begin{equation}
M^{2}-2Q^{2}=
M^{2}-({\cal Q}_{e}^{1})^{2}-({\cal Q}_{m}^{2})^{2}\geq 0\, ,
\end{equation}

\noindent with the equality satisfied in the extreme $r_{0}=0$ limit.
Performing the $\tau$ transformation on the above family of solutions
we get the dual family of solutions

\begin{equation}
\left\{
\begin{array}{rcl}
 d\tilde{\hat{s}}_{E}^{2} & = & -H^{-1}K^{-1}Wdt^{2}
+HK\left[ W^{-1}dr^{2} + r^{2}d\Omega^{2} \right] \, , \\
& & \\
e^{\tilde{\hat{\phi}}} & = &H/K\, ,\\
& & \\
\tilde{\hat{A}}^{1}{}_{t} & = & 2\alpha_{1}\frac{|Q|}{M-r_{0}}
                              \left(K^{-1}-1\right)\, ,\\
& & \\
\tilde{\hat{A}}^{2}_{\varphi} & = & -2\alpha_{2}|Q|\cos \theta\, ,
\end{array}
\right.
\end{equation}

\noindent where

\begin{equation}
K = 1 - \frac{M+r_{0}}{r}\, ,
\end{equation}

The above metric has several singularities: there is a curvature
singularity at $r=0$ and the would-be horizon singularity at $r=2r_{0}$
but both lie beyond another physical singularity at $r=M+r_{0}\geq
2r_{0}$ which is where the function $K$ vanishes and where 2-spheres of
radius $r$ have zero area. This is, therefore, a naked singularity.

Now the mass of the dual solution is clearly equal to the dilaton charge
of the original RN solution $\tilde{M}={\cal Q}_{d}=0$ and vice-versa
$\tilde{\cal Q}_{d}=M$. The electric and magnetic charges have the same
values.

This is a non-extreme massless ``black hole'' where the non-extremality
is provided by primary scalar hair.

Now, if one takes the ``extreme limit'' $r_{0}=0$ (that is, the extreme
limit in the original solution) which is also the limit in which all the
primary scalar hair vanishes and all the dilaton charge is completely
determined by the electric and magnetic charges\footnote{The situation
parallels the usual situation in which there is unconstrained ``primary
mass'' and ``secondary mass'' which is completely fixed by the electric
and magnetic charges through the B bound.} $\tilde{\cal Q}_{d}^{2}=2{\cal
Q}^{2}$ so the B bound is saturated

\begin{equation} 
\left\{
\begin{array}{rcl}
d\tilde{\hat{s}}_{E}^{2} & = & -H^{-1}K^{-1}dt^{2}
+HK\left[dr^{2} + r^{2}d\Omega^{2} \right] \, , \\
& & \\
e^{\tilde{\hat{\phi}}} & = & H/K\, ,\\
& & \\
\tilde{\hat{A}}^{1}{}_{t} & = & -\sqrt{2} \alpha_{1}
\left(K^{-1} -1\right)\, , \\
& & \\
\tilde{\hat{A}}^{2}_{\varphi} & = & -2\alpha_{2}|Q|\cos \theta\, ,
\end{array}
\right.
\end{equation}

\noindent which is  one of the  extreme massless black holes in 
Refs.~\cite{kn:BKL}, identified as composite objects in the sense of
Ref.~\cite{kn:R} in Ref.~\cite{ort2} and further studied in
Refs.~\cite{kn:E}.

Observe that, while primary scalar hair should be included in the B
bound, the primary scalar hair completely disappears in the saturated B
bound. Thus, unbroken supersymmetry acts as a {\it cosmic hairdresser} and
it is not possible to find solutions with unbroken supersymmetry and
primary scalar hair.

As a last example we consier the well-known Kerr spacetime metric
which in Boyer-Lindquist coordinates reads:

\begin{eqnarray}
d\hat{s}_{E}^{2} & = & 
-\frac{r^2 - 2Mr + a^2}{r^2 + a^2 \cos^2 \theta}(dt 
-a \sin^2 \theta d\phi)^2 
\nonumber \\
& & \nonumber \\
& &
+\frac{\sin^2 \theta}{r^2 + a^2 \cos^2 \theta}
\left[(r^2 + a^2)d\phi - adt\right]^2 \nonumber\\
& & \nonumber \\
&&
{}+\frac{r^2 + a^2 \cos^2 \theta}{r^2 - 2Mr + a^2}dr^2 
+(r^2 + a^2 \cos^2 \theta )d\theta ^2\, , 
\end{eqnarray}

\noindent where $a=J/M$. This metric belongs to a more general
class of metrics which can be written in appropriate coordinates as:

\begin{equation}
d\hat{s}_{E}^{2} = -G(dt - \omega d\phi)^2 +A dr^2 +B d\theta^2 +C d\phi^2\, ,
\end{equation}

\noindent where $G,\omega,A,B$ and $C$ are arbitrary functions of $r$ and
$\theta$, conveying the adapted character of the coordinates employed.

The T~dual with respect to the isometry with Killing vector
$\frac{\partial}{\partial t}$ is easily found to be, in the Einstein
frame,

\begin{equation}
d \tilde{\hat{s}}_{E}^{2} = -G^{-1} dt^2 
+A dr^2 +B d \theta^2 +C d \phi^2\, ,
\end{equation}

\noindent There is also a two-form present, given by

\begin{equation}
\hat{B} = - \omega d t \wedge d \phi\, ,
\end{equation}

\noindent as well as a dilaton, namely

\begin{equation}
\hat{\phi} =-{\textstyle\frac{1}{2}} \log|G |\, ,
\end{equation}

It is well known that in the static Schwarzschild case \cite{quev}
what appear as horizons in one metric, look as singularities in the
T~dual of it.  In the more general, stationary case considered here,
there are two related concepts: the infinite redshift surface, (also
called the ``static limit'') that is, the stationary limit surface
bordering the region in which the Killing ${\partial\over \partial t}$
is timelike; and the event horizon; that is the hypersurface where $r
= constant$ becomes null ; the region between those two surfaces being
the ergosphere.

In the $\tilde{Kerr}$ metric presented above, there is no ``infinite
redshift surface'', and before the surface $r=const$ becomes null, a
singularity develops, located at 

\begin{equation}
G \equiv {r^2 + a^2 \cos^2 \theta -2mr
\over r^2 + a^2 \cos^2 \theta} = 0\, .
\end{equation}

\noindent The metric is easily seen to be asymptotically flat, and the 2-form
goes to zero at infinity as

\begin{equation}
\hat{B} =  2ma \sin^2 \theta\ {1\over r} \left[ 1 + {2m\over r}
+ {\cal O}(r^{-2})\right]dt\wedge d\phi\, .
\end{equation}

%%%%%%%%%%%%%%%%%%%%%%%%%%%%%%%%%%%%%%%%%%%%%%%%%%%%%%%%%%%%%%%%%%%%%

\section{Conclusions}
\label{sec-conclusions}

The results of the present paper concerning the transformation of the
charges under duality leave unanswered the question posed in the
Introduction: why the angular momentum appears in the definition of
extremality (defining the borderline between regular horizon and a
naked singularity, with zero Hawking temperature) but not in the
Bogomol'nyi bound (whose saturation guarantees absence of quantum
corrections, as well as a ``zero force condition'', allowing
superposition of static solutions).

We now believe this is due to the fact that stationary (as opposed
to static) black holes possess a specific decay width, which can even
be seen classically by scattering waves off the black hole. This
process is known as ``superradiance'' (\cite{bdw}; see also
\cite{kn:FHM}) in the black hole literature.

The way this appears is that the amplitude for reflected waves is
greater than the corresponding incident amplitude, for low
frequencies, up to a given frequency cutoff, $m\Omega_H$, depending on
the angular momentum of the hole, and such that $\Omega_H(a=0)=0$.
The angular momentum of the hole decreases by this mechanism until a
static configuration is reached.  The physics underlying this process
is similar to the one supporting Penrose's energy extraction
mechanism, namely, the fact that energy can be negative in the
ergosphere. This, in turn, is an straightforward consequence of the
mathematical fact that the Energy of a test particle is defined as $E
= p.k$, where $p$ is the momentum of the particle, and $k$ is the
Killing vector (which has spacelike character precisely in the
ergosphere); and the product of a spacelike vector with a timelike one
does not have a definite sign.

Quantum mechanically, this means that there are two competing mechanisms
of decay for a rotating (stationary) black hole:
spontaneous emission (the quantum effect associated to the superradiance),
which is not thermal (and disappears when the angular momentum goes to zero)
 and Hawking radiation, which is thermal.

The first one is most efficient for massive black holes, but its width is
never zero even for small masses, until the black hole has lost all its
angular momentum.

This clearly shows that even if the black hole is extremal, it cannot
be stable quantum mechanically as long as its angular momentum is
different from zero. This argument taken literally would suggest that
it is not possible to have BPS states with non zero angular momentum,
unless they are such that no ergosphere exists. This is the case of
the supersymmetric Kerr-Newman solutions which are singular and,
therefore, do not have ergosphere. What is not clear is why
supersymmetry signals as special that singular case and not the usual
extremal Kerr-Newman black hole\footnote{This argument seems to be
  valid only in four dimensions, though, since rotating charged black
  holes which are BPS states exist in five dimensions \cite{kn:BMPV}.
  The existence of two Casmirs for the five-dimensional angular
  momentum seems to play an important role.}.

It could well be that Supergravity is not capable to give that answer
but String Theory is: from the String Theory point of view, given an
extreme Reissner-Nordstr\"om black hole, if we want to add angular
momentum, we can only do it at the expense of adding mass at the same
time. Thus, according to the String Theory black-hole building rules,
one can get extreme Kerr-Newman black holes but never a
supersymmetric (singular) object with non-zero angular momentum. In
this sense, while Supergravity acts as a cosmic censor only in static
cases, String Theory seems to act as a true cosmic censor in all
cases. The singular solutions cannot be built in the theory.

A similar argument could also be enough to prove a no-hair theorem in
String Theory: it could happen that it is impossible to build String
Theory states with primary scalar hair because there is no primary
source for scalar hair in it. In this sense String Theory would act as
a {\it cosmic hairdresser}. Here the situation is, though, a bit
different. First of all, we are clearly a long way from proving that
there are no microscopic configurations in String Theory that result
in macroscopic primary scalar hair. In fact, the situation resembles a
bit the situation of the ``primary mass'' (the mass that exceeds the
Bogomol'nyi identity) since it is not clear what the microscopic
configuration that manifests itself as that primary mass is and, thus,
there is no String Theory model for the Schwarzschild black hole.  It
is, in fact, conceivable that both quantities have the similar
origins, as T~duality seems to be indicating. This would be a more
attractive scenario since then we would have a tool ins String Theory
to understand no-hair theorems from first principles.

There is, yet, another, more speculative, possibility that we could
like to mention. Since extreme and non-extreme massless ``black
holes'' seem to have the same kind of singularities as their regular
T~dual counterparts (null and spacelike, respectively) one could, in
principle, use the spacetime of the massless black hole to patch up
the spacetime of the massive one, gluing them at the singularity. This
would be a non-analytic continuation through the singularity with the
help of T~duality much in the same spirit as T~duality at finite
temperature can relate high and low-temperature regimes of the
heterotic string even though, in between the free energy diverges at
the Hagedorn temperature \cite{kn:SAO}.

From the point of view of String Theory this possibility looks more
plausible when one takes into account the lower sensitivity of strings
to spacetime singularities, as compared to point particles
\cite{kn:HS},

Work in this direction is in progress.

%%%%%%%%%%%%%%%%%%%%%%%%%%%%%%%%%%%%%%%%%%%%%%%%%%%%%%%%%%%%%%%%%%%%%

\section*{Acknowledgments}

We gratefully acknowledge most useful conversations with G.W.~Gibbons
and A.A.~Kehagias.  The work of of E.A.~has been partially supported
by grants AEN/96/1655 and AEN/96/1664.The work of P.M.~has been
partially supported by the E.U.~under contract number ERBFMBICT960616.
T.O.~would like to thank M.M.~Fern\'andez for her support.

%%%%%%%%%%%%%%%%%%%%%%%%%%%%%%%%%%%%%%%%%%%%%%%%%%%%%%%%%%%%%%%%%%%%%

\end{document}